\documentclass[12pt,letterpaper]{article}
\pdfoutput=1
\usepackage{graphicx}
\usepackage{subfig}
\usepackage{amsfonts,amssymb}
\newcommand{\eqn}[1]{eq.(\ref{#1})}
\newcommand{\eqns}[2]{eqns.(\ref{#1},\ref{#2})}
\newcommand{\be}{\begin{equation}}
\newcommand{\ee}{\end{equation}}
\newcommand{\ben}{\begin{displaymath}}
\newcommand{\een}{\end{displaymath}}
\newcommand{\bea}{\begin{eqnarray}}
\newcommand{\eea}{\end{eqnarray}}
\newcommand{\bean}{\begin{eqnarray*}}
\newcommand{\eean}{\end{eqnarray*}}

\def\s {\sigma}

\renewcommand{\t}{\theta}

\newcommand{\cM}{\mbox{${\cal M}$}}

\newcommand{\ads}[1]{\mbox{${AdS}_{#1}$}}

\newcommand{\eg}{{\it e.g.}}
\newcommand{\ie}{{\it i.e.}}

\newcommand{\commentout}[1]{}

\newcommand{\beq}{\begin{equation}}
\newcommand{\eeq}{\end{equation}}
\newcommand{\beqr}{\begin{displaymath}}
\newcommand{\eeqr}{\end{displaymath}}
\newcommand{\beqa}{\begin{eqnarray}}
\newcommand{\eeqa}{\end{eqnarray}}
\newcommand{\beqar}{\begin{eqnarray*}}
\newcommand{\eeqar}{\end{eqnarray*}}

\newcommand{\cO}{{\cal O}}

\newcommand{\non}{\nonumber}

\newcommand{\half}{\ensuremath{\frac{1}{2}}}

\newcommand{\bz}{\ensuremath{\bar{z}}}

\newcommand{\ps}{\ensuremath{\partial_\sigma}}

\newcommand{\that}{\hat{\theta}}
\newcommand{\inte}[2]{{ \int_{{#1}}^{{#2}}}}
\newcommand{\X}{\mathrm{X}}
\newcommand{\zs}{\zeta_s}
\newcommand{\sn}{\mathrm{\bf sn}}

\newcommand{\dn}{\mathrm{\bf dn}}

\begin{document}
%%%%%%%%%%%%%%%%%%%%%%%%%%%%%%%%%%%%%%%%%%%%%%%%%%%%%%%%%%%%%%%%%%%%%%%%
%%%%%%%%%%%%%%%%%%%%%%%%%%%%%%%%%%%%%%%%%%%%%%%%%%%%%%%%%%%%%%%%%%%%%%%%
%%%%%%%%%%%%%%%%%%%%%% TITLEPAGE %%%%%%%%%%%%%%%%%%%%%%%%%%%%%%%%%%%%%%%
%%%%%%%%%%%%%%%%%%%%%%%%%%%%%%%%%%%%%%%%%%%%%%%%%%%%%%%%%%%%%%%%%%%%%%%%
%%%%%%%%%%%%%%%%%%%%%%%%%%%%%%%%%%%%%%%%%%%%%%%%%%%%%%%%%%%%%%%%%%%%%%%%

\title{\LARGE \bf Wilson loops and Riemann theta functions II}

\author{Martin Kruczenski\thanks{E-mail: \texttt{markru@purdue.edu}}\\
        Department of Physics, Purdue University,  \\
        525 Northwestern Avenue, W. Lafayette, IN 47907-2036,  \vspace{0.5cm} \\
        {\em and} \vspace{0.5cm}\\
         Sannah Ziama\thanks{E-mail: \texttt{spziama@pa.uky.edu}}\\
                 Department of Physics and Astronomy, University of Kentucky,  \\
                505 Rose Street, Lexington, KY 40506-0055 }

\maketitle

\begin{abstract}

 In this paper we extend and simplify previous results regarding the computation of Euclidean Wilson loops in the context of the AdS/CFT correspondence,
or, equivalently, the problem of finding minimal area surfaces in hyperbolic space (Euclidean \ads{3}). If the Wilson loop is given by a boundary 
curve $\vec{X}(s)$ we define, using the integrable properties of the system, a family of curves $\vec{X}(\lambda,s)$ depending on a complex parameter $\lambda$ known as the spectral parameter. This family has remarkable properties. As a function of $\lambda$,  $\vec{X}(\lambda,s)$ has cuts and therefore 
is appropriately defined on a hyperelliptic Riemann surface, namely it determines the spectral curve of the problem. Moreover, $\vec{X}(\lambda,s)$ has an
essential singularity at the origin $\lambda=0$. The coefficients of the expansion of $\vec{X}(\lambda,s)$ around $\lambda=0$, when appropriately integrated
along the curve give the area of the corresponding minimal area surface.

 Furthermore we show that the same construction allows the computation of certain surfaces with one or more boundaries corresponding to Wilson loop
correlators. We extend the area formula for that case and give some concrete examples. As the main example we consider a surface ending on two concentric circles 
and show how the boundary circles can be deformed by introducing extra cuts in the spectral curve.

\end{abstract}

\clearpage
\newpage

%\keywords{Classical string solutions, AdS/CFT, Wilson loops}

%\preprint{\tt{} \\
%          \tt{hep-th/yymmnnn}  }

%%%% INTRODUCTION

\section{Introduction}

 The AdS/CFT correspondence \cite{malda} set on firm footing the conjectured relation between string theory and gauge theories in the planar limit. 
 It opened the possibility of computing gauge theory quantities in the strong coupling limit by using the dual string theory. 
 Perhaps the most important observable in gauge theories is the Wilson loop. In AdS/CFT it is related to surfaces ending on the curve associated 
 to the Wilson loop \cite{MRY,DGO}. In the large 't Hooft coupling limit, the surface that needs to be considered is the one of minimal area. 
 Among all Wilson loops the simplest ones are those determined by a two dimensional, closed, smooth curve. 
 Until recently, the only such example whose dual minimal area surface could be computed was the circle \cite{cWL}. The minimal area surface was simply 
 given by a half sphere. This was somewhat surprising given the known integrability properties of the system \cite{Tseytlin:2010jv,Pohlmeyer}.
 
 Recent progress in \cite{IKZ} provided a new, infinite parameter family of minimal area surfaces that can be used to further
 explore the AdS/CFT duality. In that work the minimal area surfaces were constructed analytically in terms of Riemann theta functions associated to 
 hyperelliptic Riemman surfaces closely following previous work by M. Babich and A. Bobenko \cite{BB,BBook}. It also follows related
 work where  Wilson loops were studied or theta functions were used in similar problems, for example in \cite{JJ,Jevicki:2007aa,spiky,DV,SS,DeVega:1992xc,Larsen:1996gn,lens,Zarembo, DF,cusp,AM,AM2,scatampl,Hoare:2012nx,Irrgang:2009uj,spinchain,Leo,DK}. 
 Much of that work was motivated by the relation to scattering amplitudes \cite{AM,AM2} which we do not pursue here. A closely related approach 
can also be found in \cite{Janik} where integrability properties are applied to the computation of Wilson loops as minimal area surfaces. 
More recent developments can be found in 
\cite{Fiol:2013hna,Muller:2013rta,Ryang:2012uf,Dekel,Irrgang:2012xb,Forini:2012bb,Papathanasiou:2012hx,Drukker:2011za,Burrington:2011eh,Alday:2011pf,Kalousios:2011hc}.

 Here we continue the study of such solutions and show that, in particular, they also provide an infinite parameter family of examples of surfaces ending on multiple curves. The case we concentrate on is the one of concentric curves. The main technical difference with the previous case is that a certain periodicity condition is required, which implies a restriction on the position of the cuts and on the values of the spectral parameter. Since it is only
 one condition, it still implies an infinite parameter family of solutions. 
 
 The paper is organized as follows: in the next section we review briefly the previous results. Since the paper is essentially a continuation of 
 the previous one we do not attempt to make this paper self-contained and should be read in conjunction with \cite{IKZ}. On the other hand we give new
 expressions for the boundary curve and the area that were derived by simplifying the ones in \cite{IKZ}.
 After that, we present the method to find minimal area surfaces ending on multiple curves and give a few examples with plots of the corresponding surfaces and boundaries. In the following section we compute the area for the case of multiple curves in complete parallel with the result for a single curve.
 Finally, as illustration,  we give an example of a surface ending in eight different curves and give our conclusion in the last section.
 
 \section{Single curve case revisited}
 
 In \cite{IKZ} it was shown, based on \cite{BB,BBook} how to find an infinite parameter family of minimal area surfaces ending on a curve at the boundary. 
The method uses integrability properties of the equations to solve them in terms of Riemann theta functions. The technique is common to many integrable non-linear differential equations and is well described in several references, for example, in the book \cite{BBook}. 

 In this section we briefly review those results and derive a simplified expression for the boundary curve which allows the explicit computation 
of other important quantities. For example, a simple expression is found for the Schwarzian derivative of the boundary curve. Finally, a remarkable
formula for the area of the curve is found in terms of the behavior of the boundary curve near singular values of the spectral parameter. 

 To be concrete, we look for minimal area surfaces in hyperbolic space (EAdS$_3$) which can be parameterized by coordinates $(X,Y,Z>0)$ with a metric
\beq
ds^2 = \frac{dX^2+dY^2+dZ^2}{Z^2}= \frac{dZ^2+d\bar{\X}d\X}{Z^2}, \ \ \ \ \X=X+iY
\label{m1}
\eeq  
Given a contour $(X(s),Y(s),Z=0)\equiv(\X(s)=X(s)+i Y(s),Z=0)$ at the boundary $Z=0$ we seek a surface of minimal area ending on it and parameterized as
\beq
 Z(z,\bz),  \ \ X+iY=\X(z,\bz)
 \label{m2}
\eeq
where the coordinates $(z,\bz)$ parameterize the complex plane. In that plane we find a closed curve $z(s)$ such 
that $Z(z(s),\bz(s))=0$. Such curve maps to the contour in the boundary of (EAdS$_3$) space, and the interior maps to a minimal area surface. 
 At this time there is no generic solution to such problem of finding the minimal area surface for an arbitrary boundary. However we can give 
an infinite parameter family of minimal area surfaces and the corresponding contours where they end. It is an open problem if such solutions 
approximate all contours (and if so, in which sense). 

 The construction starts by choosing an auxiliary hyperelliptic Riemann surface $\cM$, known as the spectral curve and given, as a subspace 
of $\mathbb{C}^2$  by
\beq
 \mu^2 = \lambda \prod_{j=1}^{2g}(\lambda-\lambda_j)
 \label{m3}
\eeq
where $(\mu,\lambda)$ parameterize $\mathbb{C}^2$ and $\lambda_i\neq \lambda_j$ if $i\neq j$. There is a reality condition that requires the set of branch points $\{0,\infty,\lambda_{j=1\ldots 2g}\}$ to be symmetric under the involution $T:\lambda\leftrightarrow -1/\bar{\lambda}$.  
 A basis of one-cycles $a_i,b_i$ is chosen such that the non-trivial intersections are $a_i\circ b_j=\delta_{ij}$ and that under the involution behave as
 \beq
  T a_i = - T_{ij} a_j , \ \ \ T b_i = T_{ij} b_j 
  \label{m4}
 \eeq
 where $T_{ij}$ is a $g\times g$ symmetric matrix such that $T^2=1$. 
 
 As a matter of notation, a generic point in the Riemann surface $\cM$ is denoted as $p_i=(\mu_{p_i},\lambda_{p_i})\in\cM$. The coordinate $\lambda_{p_i}$
is called the projection of $p_i$ on the complex plane. The origin and infinity play an important role and are denoted as $p_1=0$, $p_3=\infty$. Except for the branch points, given a point $p_4$ 
in one sheet there is another point denoted as $p_{\bar{4}}$ on the other sheet of the Riemann surface such that $\lambda_{p_4}=\lambda_{p_{\bar{4}}}$ and $\mu_{p_4}=-\mu_{p_{\bar{4}}}$. The relation 
between $p_4$ and $p_{\bar{4}}$ is the hyperelliptic involution associated with $\cM$.
 
  Now we consider the basis of holomorphic differentials 
\beq
 \nu_j = \frac{\lambda^{j-1}}{\mu(\lambda)} d\lambda, \ \ \ \ \ i,j=1\ldots g
 \label{m5c}
\eeq
and define the matrices
\beq
 C_{ij} = \oint_{a_i} \nu_j, \ \ \ \ \tilde{C}_{ij} = \oint_{b_i} \nu_j
  \label{m6c}
\eeq
and the basis of normalized holomorphic differentials $\omega_{i=1\cdots g}$
\beq
 \omega_i = \nu_j (C^{-1})_{ji} = \sum_{j=1}^g C^{-1}_{ji} \lambda^{j-1} \frac{d\lambda}{\mu(\lambda)} 
  \label{m7c}
\eeq
 such that 
\beq
 \oint_{a_i} \omega_j = \delta_{ij}, 
 \label{m5}
\eeq
which defines the periodicity matrix
\beq
\oint_{b_i} \omega_j = \Pi_{ij}, 
\label{m6}
\eeq
 and the theta function 
\beq
 \theta(\zeta) = \sum_{n\in \mathbb{Z}^g} e^{i\pi\, n^t\Pi n +2\pi i\, n^t \zeta}, \ \ \ \ \zeta\in \mathbb{C}^g
 \label{m7}
\eeq
 It also defines the Jacobi map $\phi: \cM \rightarrow \mathbb{C}^g$ for which the standard notation is
\beq
 \phi(p_4) = \inte{p_1}{p_4} \omega = \inte{1}{4}
 \label{m8}
\eeq
 where the point $p_1=0$ is chosen as a distinguished point, $p_4$ is an arbitrary point and $\inte{1}{4}$ is just notation for the same function.  
There is an ambiguity in choosing the path of integration. Since two different paths from $p_1$ to $p_4$ can only differ by a closed cycle, $\phi(p_4)$
changes by
\beq
 \phi(p_4) = \inte{1}{4} \ \ \rightarrow \ \ \ \inte{1}{4} + \epsilon_2 + \Pi \epsilon_1
 \label{m9}
\eeq
 where $\epsilon_1$ and $\epsilon_2$ are vectors with integer components. Further, if $p_4$ is another branch point, the path $1\rightarrow 4$ 
 can be traced back on the lower sheet defining a generically non-trivial cycle. By choosing a certain path from $p_1=0$ to $p_3=\infty$ we define
the vectors $\Delta_{1,2}\in \mathbb{Z}^g$ through
\beq
  \inte{1}{3} = \half \Delta_2 + \half \Pi \Delta_1 
  \label{m10}
\eeq
 The path should be chosen such that $\Delta^t_1.\Delta_2$ is an odd integer. With this vector we define another function
\beq
 \that(\zeta) = \exp\left\{2\pi i\left[\frac{1}{8}\Delta_1^t\Pi\Delta_1+\half\Delta_1^t\zeta+\frac{1}{4}\Delta_1^t\Delta_2\right] \right\}
 \theta(\zeta+\inte{1}{3})  
\label{m11}
\eeq
 Because   $\Delta^t_1.\Delta_2$ is odd it follows that $\that(-\zeta)=-\that(\zeta)$ and in particular $\that(0)=0$.
 
Further, consider the vector $\omega=(\omega_1,\cdots,\omega_g)$ and expand it around $p_1=0$. Since $p_1=0$ is a branch point an appropriate coordinate
can be chosen as $y =-2i\sqrt{\lambda}$, namely $\lambda=-y^2/4$. It follows that
\beq
 \omega = (\omega_1 + y^2 \omega_{12} + y^4 \omega_{14} +\cdots ) dy
 \label{m12}
\eeq
for some constant vectors $\omega_1$, $\omega_{12}$, etc. that can be computed from \eqn{m7c} by expanding $\mu(\lambda)$.
 Near infinity, on the other hand, an appropriate coordinate is $\tilde{y}=\frac{2}{\sqrt{\lambda}}$ and the expansion is 
\beq
 \omega = (\omega_3 + \tilde{y}^2 \omega_{32} + \tilde{y}^4 \omega_{34} +\cdots) d\tilde{y}
 \label{m13}
\eeq
 The vectors $\omega_1, \omega_3, \omega_{12}, \omega_{32},\ldots \in\mathbb{C}^g$ play an important role and it is convenient to introduce a notation 
for the gradient of a function along those directions 
\beqa
  D_1 F(\zeta) &=& (\omega_1 . \nabla) F(\zeta) \\
  D''_1 F(\zeta) &=& (\omega_{12} . \nabla) F(\zeta) \\
  D_3 F(\zeta) &=& (\omega_3 . \nabla) F(\zeta) \\
  D''_3 F(\zeta) &=& (\omega_{32} . \nabla) F(\zeta) 
   \label{m14}
\eeqa
Notice that $D''_1$ is a first derivative. Second derivatives are denoted \eg\ as $D_1^2 F(\zeta) = (\omega_1 . \nabla) (\omega_1 . \nabla) F(\zeta)$.
With all these ingredients in place, the solutions are given by
 \beqa
  Z&=& \left|\frac{\that(2\inte{1}{4})}{\that(\inte{1}{4})\theta(\inte{1}{4})}\right|
     \frac{|\theta(0)\theta(\zeta)\that(\zeta)||e^{\mu z+\nu \bz}|^2}{|\that(\zeta-\inte{1}{4})|^2+|\theta(\zeta-\inte{1}{4})|^2} \label{m15a} \\
 X+iY &=& \X= e^{2\mu z+2\nu \bz} \frac{\theta(\zeta+\inte{1}{4})\overline{\theta(\zeta-\inte{1}{4})}-\that(\zeta+\inte{1}{4})\overline{\that(\zeta-\inte{1}{4})}}{|\that(\zeta-\inte{1}{4})|^2+|\theta(\zeta-\inte{1}{4})|^2} \ \ \ \ \ 
 \label{m15}
 \eeqa
where the vector $\zeta(z,\bar{z}) \in \mathbb{C}^g$ is given by
\beq
   \zeta = 2\omega_1 \bz + 2\omega_3 z
   \label{m16}
\eeq   
and the constants $\mu$, $\nu$ are given by
\beq
 \mu = -2 D_3\ln\theta(\inte{1}{4}), \ \ \ \ \nu=-2D_1\ln\that(\inte{1}{4})
\label{munu}
\eeq
which uses the directional derivatives $D_{1,3}$ defined in \eqn{m14}. Notice that \eqn{m16} implies $\partial_z F(\zeta(z,\bz)) = 2 D_1 F(\zeta(z,\bz))$
 and $\partial_{\bz} F(\zeta(z,\bz)) = 2 D_3 F(\zeta(z,\bz))$. The solution contains an arbitrary parameter $p_4=(\mu_{p_4},\lambda_{p_4})\in\cM$.  
For \eqn{m15} to define a solution it is necessary that $|\lambda_{p_4}|=1$. Therefore, each spectral curve $\cM$ actually defines a one real parameter 
family of surfaces. It was shown in \cite{IKZ} that all those surfaces have the same area.

 Since the boundary is at $Z=0$, to find the curve where the surface ends we need to find the zeros of $Z$ or, equivalently, the zeros of $\that$. 
Typically the zeros are given by isolated curves $z(s)$ on the world-sheet from which one can be chosen. 
The solution \eqn{m15} maps such curve $z(s)$ to a curve $\X(s)$ at the boundary $Z=0$ defining the shape of the Wilson loop. 
The region of the world-sheet inside the curve $z(s)$ is mapped, by \eqn{m15} to the minimal area surface ending at the curve $\X(s)$.
 
 \subsection{Boundary curve}
 
 As just described the shape of the boundary curve, or equivalently of the dual Wilson loop, follows from finding a curve $z(s)$ on the world sheet 
such that
 \beq
  \hat{\theta}(\zs) =0 
  \label{m17}
\eeq
where
\beq
\zs = 2 \omega_1 \bar{z}(s) + 2 \omega_3 z(s)
\label{m18}
\eeq
 Once this function $\zs$ is found, the shape of the curve follows from replacing $\zeta\rightarrow\zs$ in the expression for $X$, namely eq.(\ref{m15}) to get
$\X(s)=X(s)+iY(s)$ \cite{IKZ}:
\beq
\X(s) = e^{2\mu z(s)+2\nu \bz(s)}
  \frac{\theta(\zs+\inte{1}{4})\overline{\theta(\zs-\inte{1}{4})}-\that(\zs+\inte{1}{4})\overline{\that(\zs-\inte{1}{4})}}
       {|\that(\zs-\inte{1}{4})|^2+|\theta(\zs-\inte{1}{4})|^2} 
\label{m19}
\eeq
  However, it turns out that the result can be simplified by using \eqn{b8} derived in the appendix. With this identity, \eqn{m19} reduces to
\beq
 X(s)+iY(s) = \X(s) = -e^{2\mu z(s)+2\nu \bz(s)}\frac{\that(\zs+\inte{1}{4})}{\that(\zs-\inte{1}{4})}
 \label{m21}
\eeq
 which is a much simpler expression for the boundary curve. In fact the function $\X(s)$ describes a family of curves parameterized by the point $p_4$
in the Riemann surface. Now we study the analytic properties of $\X(s,p_4)$ as a function of $p_4$. 

\subsection{Analytic properties of $\X(s,p_4)$, analogy with the monodromy matrix}

 As discussed in the previous section, the function
 \beq 
 \X(s,p_4)=-e^{2\mu z(s)+2\nu \bz(s)}\frac{\that(\zs+\inte{1}{4})}{\that(\zs-\inte{1}{4})}
 \label{m22}
 \eeq
can be thought of as a family of curves parameterized by a point $p_4$ on the Riemann surface $\cM$. When $|\lambda_{p_4}|=1$ the curve is such that
the minimal area surface can be found from eqns. (\ref{m15a}) and  (\ref{m15}). In this section we study the analytic properties of $\X(s,p_4)$ as a function of $p_4$
and show that it is well defined on the Riemann surface but not on the complex plane parameterized by $\lambda_{p_4}$ where it has cuts. 
In that manner, $\X(s,p_4)$ defines the spectral curve in the same way as the monodromy matrix does in many integrable systems. 
 Furthermore, it turns out that $\X(s,p_4)$ has essential singularities at $p_4=0,\infty$, the precise behavior around the singularities determines
the area of the corresponding surface. Finally, the Schwarzian derivative $\{\X,s\}$ is computed, and shown to have a remarkable simple dependence 
on $\lambda_{p_4}$. 
 
 To understand the properties of $\X(s,p_4)$ as a function of $p_4$ first it has to be proven that it is a well defined function on the Riemann surface.  
The reason it might not be is that there is an ambiguity in choosing the path of integration defining $\inte{1}{4}$. Choosing another path changes
the integral by a period of the theta function. Using the quasi periodicity properties of $\that$ \ie\ \eqn{b2}, it follows that
 \beqa
  \frac{\that(\zs+\inte{1}{4}+\epsilon_2+\Pi\epsilon_1)}{\that(\zs-\inte{1}{4}+\epsilon_2+\Pi\epsilon_1)}&=&
  e^{-4\pi i \epsilon_1^t\zeta e^{2\pi i[\Delta_1^t\epsilon_2-\epsilon_1^t\Delta_2]}}
  \frac{\that(\zs+\inte{1}{4})}{\that(\zs-\inte{1}{4})} \\
  &=&  e^{-4\pi i \epsilon_1^t\zeta}
    \frac{\that(\zs+\inte{1}{4})}{\that(\zs-\inte{1}{4})}
    \label{m23}
 \eeqa 
 where we used that $\Delta_1^t\epsilon_2-\epsilon_1^t\Delta_2$ is an integer since $\epsilon_{1,2}$ and $\Delta_{1,2}$ are integer vectors. 
 Therefore the ratio of theta functions is {\em not} a well defined function on the Riemann surface. However, the exponential factor also depends 
on $p_4$ through the constants $\mu$, $\nu$ defined in \eqn{munu}:
\beq
 \mu = -2 D_3\ln\theta(\inte{1}{4}), \ \ \ \ \nu=-2D_1\ln\that(\inte{1}{4})
\label{m24}
\eeq
 To study their periodicity properties recall the definition
 \beq
   D_{p_i} F(\zeta) = \omega(p_i)_k \nabla_k F(\zeta)
   \label{m25}
 \eeq
and introduce a variable $\zeta$ to compute the derivatives
 \beqa
  \mu\rightarrow \tilde{\mu}&=& -2 \left. D_{p_3}\ln \theta(\zeta+\inte{1}{4}+\epsilon_2+\Pi\epsilon_1)\right|_{\zeta=0} \\
                  &=& -2\left. D_{p_3} \left[\ln\theta(\zeta+\inte{1}{4})-2\pi i \epsilon_1^t \zeta\right]\right|_{\zeta=0}     \\
                  &=& \mu +4\pi i \epsilon_1^t \omega_3
 \label{m26}
 \eeqa
 Similarly
 \beq
 \nu \rightarrow \tilde{\nu} = \nu + 4\pi i \epsilon_1^t \omega_1
 \label{m27}
 \eeq
 Therefore the exponential factor transforms as
 \beq
 e^{2\mu z(s)+2\nu \bz(s)} \rightarrow e^{2\mu z(s)+2\nu \bz(s)} e^{4\pi i \epsilon_1^t(2\omega_3 z(s)+2\omega _1\bz(s))} 
         =e^{2\mu z(s)+2\nu \bz(s)} e^{4\pi i \epsilon_1^t\zs}
 \label{m28}
 \eeq
 The factor precisely cancels the factor in \eqn{m23} coming from the ratio of theta functions showing that $\X(s,p_4)$ is the correct combination 
that is well defined on the Riemann surface.  
 
  Having proven that $\X(s,p_4)$ is well defined on the Riemann surface, it is interesting to study if it takes different values on the two sheets 
of the Riemann surface. This is indeed so. Changing sheets on the Riemann surface is equivalent to change $4\rightarrow \bar{4}$ or equivalently
 \beq
 \inte{1}{4} \rightarrow -\inte{1}{4}
\label{m29}
 \eeq
 Since $\theta$ is even and $\that$ is odd, the functions $\mu$ and $\nu$ change sign under such change. The ratio of theta functions is obviously 
inverted resulting in:
 \beq
  \X(s,p_{\bar{4}}) = \frac{1}{\X(s,p_4)} 
 \label{m30}
\eeq
 Therefore the function $\X(s,\lambda_{p_4})$ defined on the complex plane has cuts. Thus, in some sense, it is analogous to the monodromy matrix.  
Namely,  the cuts in $\X(s,\lambda_{p_4})$ can be removed by extending the function to $\X(s,p_4)$ defined on a Riemann surface, the spectral curve of the problem. 

Let us now study the analytical properties of $\X(s,p_4)$. Riemann's theorem \cite{FK} ensures that theta functions defined on a Riemann surface 
have $g$ zeros and no poles. The ratio that appears in the definition of $\X(s,\lambda_{p_4})$ therefore can have at most $g$ zeros and $g$ poles. 
One however is at $p_4=p_1$ and therefore cancels. In fact for the others it is clear that if $p_4$ is a zero then $p_{\bar{4}}$ is a pole. 
So $\X(s,p_4)$ has $(g-1)$ zeros and $(g-1)$ poles conjugate from each other on the Riemann surface (that means on opposite sheets).
On the other hand the exponential function is quite interesting. The function 
\beq
 \mu=-2 D_3\ln \theta(\inte{1}{4}) = -2\frac{D_3\theta(\inte{1}{4})}{\theta(\inte{1}{4})}
 \label{m31}
\eeq
 has potential singularities on the $g$ zeros of $\theta(\inte{1}{4})$. However, the derivative also vanishes on those zeros except at $p_4=p_1$. 
Since $\mu$ has a pole at $p_4=p_1=0$ then $\X(s,p_4)$ has an essential singularity. The same happens at $p_4=p_3$ in view of the function $\nu$.

 To summarize: $\X(s,p_4)$ as a function of $p_4$ is a well defined function on the Riemann surface with $(g-1)$ zeros and $(g-1)$ poles and essential
singularities at $p_4=p_1=0$ and $p_4=p_3=\infty$. 

Given the symmetry $\lambda\leftrightarrow-1/\bar{\lambda}$ it suffices to study the function  
near $\lambda=0$. Since $\lambda=0$ is a branch point a more appropriate coordinate $y$ can be chosen as
\beq
 y = - 2 i \sqrt{\lambda}
\label{m32}
\eeq
Using the expansion from \eqn{m12}
\beq
 \omega(y) = (\omega_1 + \omega_{12} y^2 + \cdots) dy
\label{m33} 
\eeq
it follows that 
\beq
\inte{1}{4} \omega = \omega_1 y + \frac{1}{3} \omega_{12} y^3 +\cdots
\label{m34}
\eeq
A simple Taylor expansion in $y$ gives
\beq
 \ln \left(-\frac{\that(\zs+\inte{1}{4})}{\that(\zs-\inte{1}{4})}\right) = 1 + y \frac{D_1^2\that(\zs)}{D_1\that(\zs)} + \cdots
 \label{m35}
\eeq
Finally the expansion 
\beqa
 \mu &=& -2 D_{13}\ln\theta(0) y + \cdots \\
 \nu &=& -3\frac{2}{y} -\frac{2}{3} y \left(\frac{D_1^3\that(0)}{D_1\that(0)}-\frac{D_1''\that(0)}{D_1\that(0)}\right) + \cdots
\label{m36}
\eeqa
and \eqn{b30a} in the appendix, result in
\beqa
 \ln \X &=& -\frac{4}{y} \bz(s) \non\\
        && + y\left[- \frac{4}{3}\left(\frac{D_1^3\that(0)}{D_1\that(0)}-\frac{D_1''\that(0)}{D_1\that(0)}\right) \bz(s)  -
 4 D_{13}\ln \theta(0) z(s)  + 2 D_1 \ln \theta(0)\right] \non\\
 && +\cO(y^3)
 \label{m38}
\eeqa
This expansion is remarkable in two respects. First, the residue of the pole $\frac{1}{y}$ contains $z(s)$, namely the parameterization of the
conformal world-sheet coordinate in terms of the boundary parameter. Second, as we will see later the constant term determines the area of the 
minimal area surface with this boundary. Actually the area of the whole family of minimal area surfaces parameterized by $p_4$ such that 
$|\lambda_{p_4}|=1$.  It is also instructive to compute the expansion of the Schwarzian derivative $\{\X,p_4\}$ based on the expansion of $\ln \X$. 
If, as shown, $\X$ has an expansion
\beq
 \X = \frac{a_1}{y} + a_2 y + \cO(y^3)
 \label{m39c}
\eeq
and since
\beq
 \{\X,y\} = \{\ln\X,y\}-\half (\partial_y \ln\X)^2 
 \label{m40c}
\eeq
 it follows that 
\beq
\{\X,y\} =-\frac{a_1^2}{2y^4} + \frac{a_1a_2}{y^2} + \cO(1) 
\label{m41c}
\eeq
 where the two leading terms come from the last term in \eqn{m40c}. It follows that
\beqa
 \{\X,y\} &=& -\frac{8}{y^4} (\bz(s))^2 \non\\
        && - \frac{4\bz(s)}{y^2}   \left[- \frac{4}{3}\left(\frac{D_1^3\that(0)}{D_1\that(0)}-\frac{D_1''\that(0)}{D_1\that(0)}\right) \bz(s)  -
 4 D_{13}\ln \theta(0) z(s)  + 2 D_1 \ln \theta(0)\right] \non\\
 && +\cO(1)
 \label{m42c}
\eeqa

\subsection{Conformal invariant ratios and Schwarzian derivative}

 In the case of the monodromy matrix one can consider its trace which is a well defined function on the complex plane since the cuts are removed. 
In the present case we can consider functions of $\X(s,p_4)$ which are invariant under inversion. Those are well defined in the complex plane. 
For example the functions
\beqa
Y^{[4]}(s_1,s_2,s_3,s_4,\lambda_{p_4})  &=& \frac{(\X(s_1,p_4) - \X(s_2,p_4))(\X(s_3,p_4) - \X(s_4,p_4))}{(\X(s_1,p_4) 
  - \X(s_3,p_4))(\X(s_4,p_4) - \X(s_2,p_4))} \non \\
Y^{[2]}(s_1,s_2,\lambda_{p_4}) &=&   \frac{\partial_{s_1} \X(s_1,p_4) \partial_{s_2} \X(s_2,p_4)}{(\X(s_1,p_4) - \X(s_2,p_4))^2}  \\
Y^{[1]}(s,\lambda_{p_4}) &=& \left\{ \X(s),s \right\} = \frac{\partial_s^3 \X(s,p_4)}{\partial_s \X(s,p_4)} -\frac{3}{2} \left(\frac{\partial_s^2 \X(s,p_4)}{\partial_s\X(s,p_4)}\right)^2 \non \\
\label{m39}
\eeqa
are invariants depending on four, two and one point of the curve respectively. In fact they are invariant under the full M\"obius group. 
The last one introduces the Schwarzian derivative which is conformally invariant. On the left hand side we emphasized that these are functions of $\lambda_{p_4}$ namely the projection 
of $p_4$ on the complex plane.  As functions of $\lambda_{p_4}$ they have no cuts. The invariant $Y^{[4]}$ can be simply written by using
\eqn{m22}. On the other hand $Y^{[2]}$ requires evaluating $\partial_s \X$. From \eqns{m22}{m18} it follows that
 \beq
 \partial_s \ln \X = 2\mu \partial_s z +2\nu \partial_s \bz + 2 \partial_s z D_1 \ln \frac{\that(\zs+\inte{1}{4})}{\that(\zs-\inte{1}{4})}
  + 2 \partial_s \bz D_3 \ln \frac{\that(\zs+\inte{1}{4})}{\that(\zs-\inte{1}{4})}
   \label{m40c}
 \eeq
 Using the definition of $\mu$ ,$\nu$ \ie\ \eqn{munu} and  \eqns{b15b}{b24} for the derivatives of the theta function we immediately find that
 \beq
  \partial_s \ln \X = -2 \partial_s z \frac{\that(2\inte{1}{4})D_3\that(0)}{\theta^2(\inte{1}{4})} \frac{\theta^2(\zeta)}{\that(\zs+\inte{1}{4})\that(\zs-\inte{1}{4})}
   \label{m41c}
\eeq  
 Equivalently
\beq
 \partial_s \X = 2 \partial_s z\, e^{2\mu z(s)+2\nu \bz(s)}\, \frac{\that(2\inte{1}{4})D_3\that(0)}{\theta^2(\inte{1}{4})}
 \frac{\theta^2(\zs)}{\that^2(\zs-\inte{1}{4})}
  \label{m42c}
\eeq
 Thus
\beqa
\lefteqn{ Y^{[2]}(s_1,s_2,\lambda_{p_4}) = \left(\frac{\that(2\inte{1}{4})D_3\that(0)}{\theta^2(\inte{1}{4})}\right)^2 \times} &&  \label{m43c}\\
&& \frac{4\partial_s z_1\partial_s z_2 \theta^2(\zeta_{s1})\theta^2(\zeta_{s2})}
  {[e^{ \mu z_{12}+\nu\bz_{12}}\that(\zeta_{s1}+\inte{1}{4})\that(\zeta_{s2}-\inte{1}{4})-
   e^{-\mu z_{12}-\nu\bz_{12}}\that(\zeta_{s1}-\inte{1}{4})\that(\zeta_{s2}+\inte{1}{4})]^2}  \non
\eeqa
with the notation $z_{12}=z_1-z_2$, $\bz_{12}=\bz_1-\bz_2$ and $z_1=z(s_1)$, $z_2=z(s_2)$.

 Finally, the Schwarzian derivative is harder to compute since it involves up to third order derivatives. However, it turns out that the
result is remarkably simple. The second derivative can be computed in a similar way as the first one resulting in 
\beq
\ps \ln \ps\X = \frac{\ps^2\X}{\ps \X}= \frac{\ps^2 z}{\ps z} + 2\ps z \left(\mu + 2D_3\ln \frac{\theta(\zeta)}{\that(\zeta-\inte{1}{4})} \right)
 \label{m44c}
\eeq
 In fact now, the Schwarzian derivative can be evaluated as
\beqa
 \{\X, s\} &=& \{z,s\} + 8(\partial_s z)^2\left[D_3^2\ln\frac{\theta(\zs)}{\that(\zs-\inte{1}{4})} 
 -\left(D_3\ln\frac{\theta(\zs)}{\theta(\inte{1}{4})\that(\zs-\inte{1}{4})}\right)^2\right] \non\\
&& - 8 (\partial_s\bz)^2 \left(\frac{D_1\that(0)\theta(\inte{1}{4})}{\theta(0)\that(\inte{1}{4})}\right)^2
 \label{m45c}
\eeqa
 The term in square brackets can be simplified by  first using \eqn{b31} to replace $D_3^2\ln\that(\zs-\inte{1}{4})$ and then \eqns{b14a}{b34} to
 further reduce the result. The second line in \eqn{m45c} is simplified by the identity (\ref{b33a}).  In that manner the result
\beq
\{\X,s\} = \{z,s\} -2 \lambda_{p_4} (\partial_s z)^2 + \frac{2}{\lambda_{p_4}} (\partial_s\bz)^2 
 - 8 (\partial_s z)^2  \left[\frac{D_3^3\that(\zs)}{D_3\that(\zs)}-3\frac{D_3^2\theta(\zs)}{\theta(\zs)}\right]
  \label{m46c}
\eeq
is obtained. The last term can be rewritten using \eqn{b32c} 
\beqa
\{\X,s\} &=& \{z,s\} -2 \lambda_{p_4} (\partial_s z)^2 + \frac{2}{\lambda_{p_4}} (\partial_s\bz)^2  \\
 && + 8 (\partial_s z)^2  \left[ 2 D^2_3\ln\theta(\zs) - \frac{D_3^3\that(0)}{D_3\that(0)}+\frac{D_3^2\theta(0)}{\theta(0)} \right]
  \label{m47c}
\eeqa
A completely equivalent expression can be found in terms of derivatives $D_1$:
\beqa
\{\X,s\} &=& \{\bz,s\} -2 \lambda_{p_4} (\partial_s z)^2 + \frac{2}{\lambda_{p_4}} (\partial_s\bz)^2  \\
 && + 8 (\partial_s \bz)^2  \left[ 2 D^2_1\ln\theta(\zs) - \frac{D_1^3\that(0)}{D_3\that(0)}+\frac{D_1^2\theta(0)}{\theta(0)} \right]
  \label{m48c}
\eeqa
 The most important point is that the dependence in $\lambda$ is very explicit in \eqn{m46c}. 
We believe this to be one of the main results of this paper since $\{\X,s\}$ gives a conformal invariant characterization of the Wilson loop.

 \subsection{Area}

\begin{figure}
\centering
\includegraphics[width=14cm]{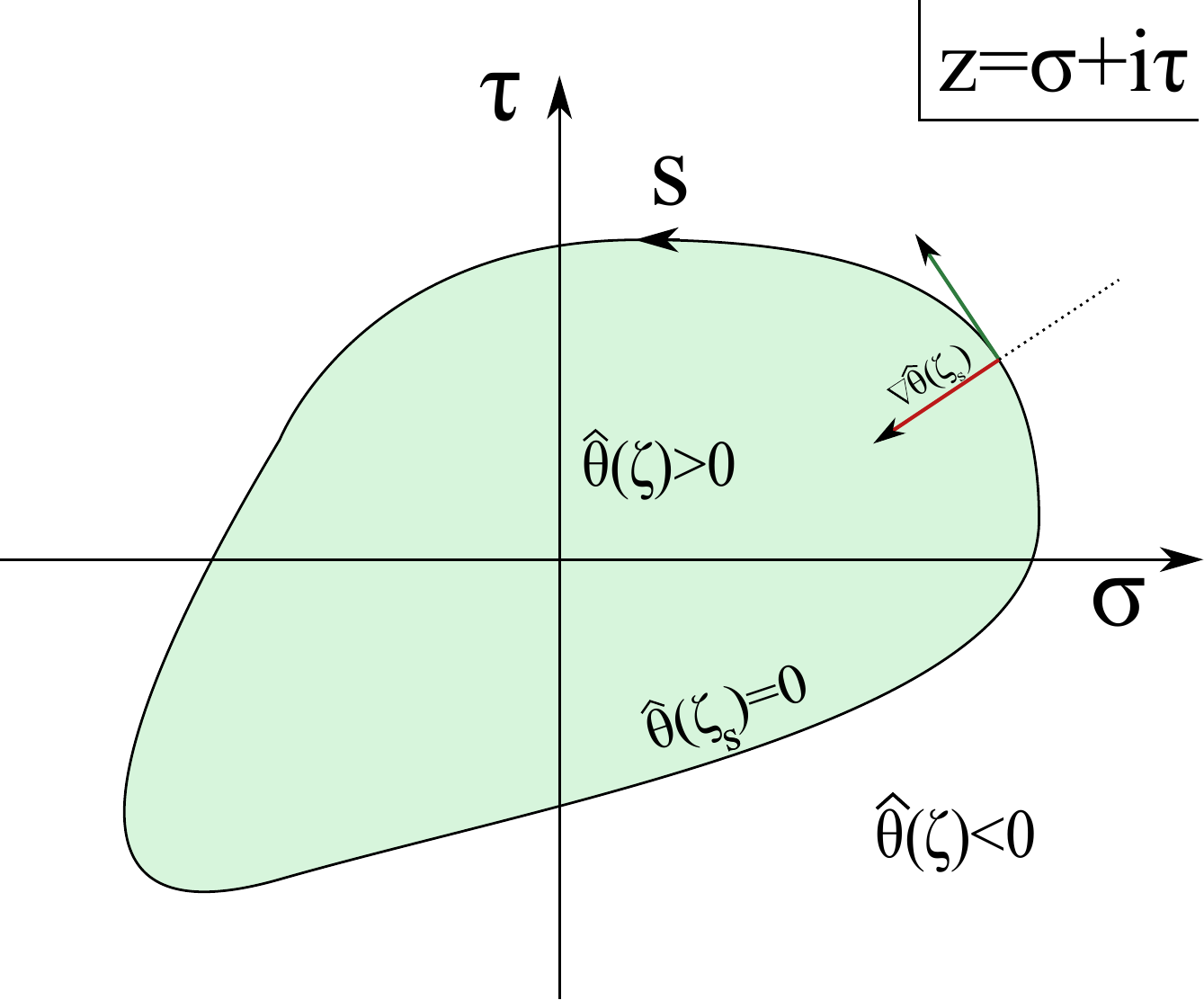}
 \caption{The world-sheet is parameterized by a complex coordinate $z$. The function $\that(\zeta)$ depends on $z$ through the 
 vector $\zeta=2\omega(p_1)\bz+2\omega(p_3)z$ where $\omega_{p_{1,3}}$ are complex vectors. 
 Our convention is that the shaded region where $\that(\zeta)>0$ maps to 
 the minimal area surface. The contour $\that(\zs)=0$ maps to the boundary of the surface, \ie\ the Wilson loop. The contour
 is parameterized counterclockwise by $s$. The gradient $\nabla \that(\zs)$ is normal to the boundary and point towards the interior. 
Finally, the region where $\that(\zs)<0$ also maps to a minimal area surface and the corresponding can be easily changed to cover that case. 
Slightly more involved is the case when $\that(\zeta)$ is purely imaginary but nothing fundamentally different changes. }
\label{fig1}
\end{figure}

 In \cite{IKZ} the following formula for the finite part of the area of the minimal area surface was derived:
\beq
   A_f = 16 D_{13} \ln \theta(0) \int d\s d\tau + \half \oint \frac{\nabla^2\that(\zs)}{|\nabla\that(\zs)} d\ell
    \label{m41}
\eeq 
 where $(\s,\tau)$ are world-sheet coordinates such that $z=\s+i\tau$ and
\beq
d\ell =\sqrt{d\s^2+d\tau^2} = \sqrt{(\partial_s\s)^2+(\partial_s\tau)^2} ds
\label{m41c}
\eeq 
 is the element of arc-length on the boundary curve $z(s)=\s(s)+i\tau(s)$.
In this section we simplify the result and relate it to the expansion of $\X(s,p_4)$ near the singularity at $p_4=0$.  
In fact, the second term in \eqn{m41} can be rewritten as
\beq
  \half \oint \frac{\nabla^2\that(\zs)}{|\nabla\that(\zs)} d\ell 
  =  \oint \partial_z\partial_{\bz} \that(\zs) \sqrt{\partial_s z \ps\bz}{\partial_z \that(\zs) \partial_{\bz} \that(\zs)} ds
   \label{m42}
\eeq
Using that $\partial_z \that = 2 D_3\that$, $\partial_{\bz} \that=2 D_1\that$ and that the contour of integration is such that $\that(\zs)=0$,
the formula for the area can be rewritten as
\beq
 A_f =  16 D_{13} \ln \theta(0) \int d\s d\t  - 2i \oint D_1\ln D_3 \that(\zs) ds
  \label{m43}
\eeq
 where the sign was chosen according to the conventions in fig. \ref{fig1}.
 
% % % % % % % % % % % % % % % % % % % % % % % % % % % % % % % % % % % % % % % % % % % % % % % % % % % % % % % % % % % % % % %
% % % % % % % % % % % % % % % % % % % % % % % % % % % % % % % % % % % % % % % % % % % % % % % % % % % % % % % % % % % % % % %
% % % % % % % % % % % % % % % % % % % % % % % % % % % % % % % % % % % % % % % % % % % % % % % % % % % % % % % % % % % % % % %
% % % % % % % % % % % % % % % % % % % % % % % % % % % % % % % % % % % % % % % % % % % % % % % % % % % % % % % % % % % % % % %

 We can further use that
\beq
 \ps \ln D_3 \that(\zs) = 2 \partial_s z D_3 \ln D_3 \that(\zs) + 2 \partial_s \bz D_1 \ln D_3 \that(\zs)
  \label{m44}
\eeq
and \eqn{b29a} to find
\beq
 A_f =  16 D_{13} \ln \theta(0) \int d\s d\tau + 4i \oint D_3 \ln \theta(\zs) dz - i \left.\ln D_3\that(\zs)\right|_{s_i}^{s_f}
  \label{m45}
\eeq
where $s_{i,f}$ are the initial and final values of $s$.  
Since the curve is periodic, 
\beq
  \left.\ln D_3\that(\zs)\right|_{s_i}^{s_f} = 2 \pi i n
   \label{m46}
\eeq
for an integer $n$ known as the turning number of the curve. 
It is the number of times the unit normal to the curve goes around the unit circle when the point goes once around the curve.
 With the assumptions of fig.\ref{fig1} it is $n=1$. The final result for the area is therefore
\beq
A_f = - 2 \pi n + 16 D_{13} \ln \theta(0) \int d\s d\tau  + 4 i \oint D_3 \ln \theta(\zs) dz 
 \label{m47}
\eeq
which is simpler than the one found in the previous paper \cite{IKZ}. 
Equivalently we can write
\beqa
A_f &=& -2\pi n - 8 i  D_{13} \ln \theta(0) \oint \bz(s) \partial_s z(s) ds + 4 i \oint D_3 \ln \theta(\zs) \partial_s z(s) ds \non \\
    &=& -2\pi n + 8 i  D_{13} \ln \theta(0) \oint z(s) \partial_s \bz(s) ds - 4 i \oint D_1 \ln \theta(\zs) \partial_s \bz(s) ds   \non \\
     \label{m48}
\eeqa
Consider now the expansion in \eqn{m38}. It immediately follows that 
\beq
 A_f = -2\pi n - \left.\left(\frac{2i}{y}\oint \ln X \partial_s \bz ds\right)\right|_{y\rightarrow 0}
  \label{m49}
\eeq
 where we recall that $y=-2i\sqrt{\lambda_{p_4}}$ is the appropriate coordinate around the branch point $p_4=0$. Also the integral of a total derivative was discarded. Quite remarkably, the function  
$\bz(s)$ is also contained in the expansion, in fact
\beq
 -\left. \frac{1}{4} y \partial_s \ln \X \right|_{y\rightarrow 0} = \partial_s \bz
  \label{m50}
\eeq
 We can then write the slightly more cumbersome formula
 \beq
  A_f = -2\pi n -\frac{2i}{\pi^2} \left(\oint \left[ \oint_{y=0} \frac{dy}{y^2} \ln X(s,y) \right] 
                                              \left[ \oint_{y'=0} dy' \partial_s\ln X(s,y') \right] ds\right)
\label{m51}
 \eeq
  The integrals on spectral parameters $y$, $y'$ are complex contour integrals (given by the residues) around the points $y=0$, $y'=0$. 
The integral along the real parameter $s$ is an integral around the boundary curve.  
 
\subsection{Genus $g=1$ case}
\begin{figure}[h!]
   \centering
     \includegraphics[width=0.9\textwidth]{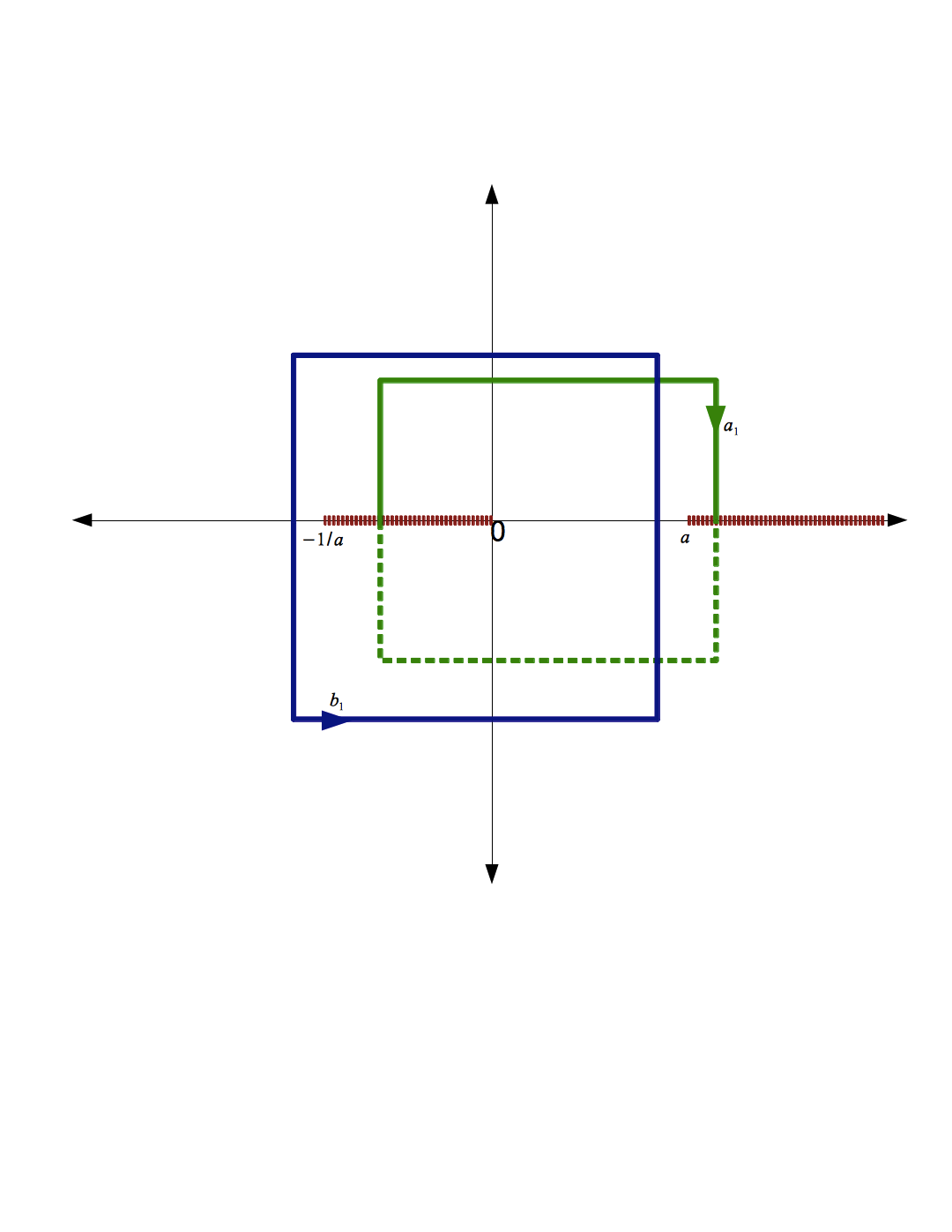}
     \vspace{-1.4in}
      \caption{\em g=1 Riemann surface.}
      \label{fig2}
 \end{figure}

 In the case of genus one the theta functions reduce to Jacobi theta functions and the solutions can be written in terms of elliptic functions or, 
equivalently, elliptic integrals. In the previous work \cite{IKZ} this case was not described since it had already been studied in \cite{Zarembo} and \cite{DF}. 
More recently, these solutions were analyzed including the evaluation of the first quantum correction \cite{DK}. 
In this subsection however, revisiting this case is useful as a starting point for the next section. For genus one there is only two cuts, namely four
branch points. They are taken to be $p_1=0$, $p_3=\infty$, $\lambda_1=a\in\mathbb{R}$, $\lambda_2=-\frac{1}{a}$ in view of the $\lambda\leftrightarrow-1/\bar{\lambda}$ symmetry. 
 The spectral curve is described by the equation
 \beq
 \mu^2 = \lambda (\lambda-a)(\lambda+\frac{1}{a})
  \label{m52}
 \eeq
an is illustrated in fig. \ref{fig2} where it is also shown a choice of cycles $a_1$, $b_1$. The only holomorphic differential is
\beq
 \nu_1 = \frac{d\lambda}{\mu(\lambda)}
\eeq
which needs to be properly normalized. Computing
 \beqa
 C_1 &=& \oint_{a_1} \frac{d\lambda}{\mu(\lambda)} = 2i \int_0^a \frac{d\lambda}{\sqrt{(a-\lambda)\lambda(\lambda+\frac{1}{a})}} 
   = 4 i \sqrt{kk'}\ \mathbf{K} \\
 \tilde{C}_1 &=& \oint_{b_1} \frac{d\lambda}{\mu(\lambda)} = - 2 \int_{-\frac{1}{a}}^0 \frac{d\lambda}{\sqrt{(a-\lambda)(-\lambda)(\lambda+\frac{1}{a})}} =
    - 4 \sqrt{kk'}\ \mathbf{K}' 
 \eeqa
where the standard notation for elliptic functions was used (see \cite{GR} )
\beq
  k = \frac{a}{\sqrt{1+a^2}}, \ \  k'=\sqrt{1-k^2}=\frac{1}{\sqrt{1+a^2}}, \ \ \ \mathbf{K} = K(k), \ \ \ \mathbf{K}' =K(k')
\eeq
The formulas from \cite{IKZ} reduce to 
\beqa
 \Pi &=& \tilde{C}_1 C^{-1}_1 = i \frac{\mathbf{K}'}{\mathbf{K}} \\
 \omega &=& -\frac{i}{4 \sqrt{kk'}\ \mathbf{K}} \ \frac{d\lambda}{\mu(\lambda)}\\ 
 \omega_1 &=& - \omega_3 = \frac{i}{4\sqrt{kk'}\, \mathbf{K}}
\eeqa
We also have $\Delta_1=1$, $\Delta_2=1$. Taking the involution $T=-1$ we find
 \beq
 \bar{\Pi} = -T\Pi T, \ \ \ \ T\Delta_{1,2}=-\Delta_{1,2}\ \ \ T\omega_3 = \bar{\omega}_1
 \eeq
as it should be for the reality condition to be satisfied. Let us introduce also the half-period $\hat{a}$ that defines $\that$ and which is a zero of $\theta$:
\beq
 \hat{a} = \frac{1+\Pi}{2}
\eeq
 Now it is possible to define the variable $\zeta$
\beq
  \zeta = 2 \omega_1 \bz + 2 \omega_3 z = \frac{\tau}{\sqrt{kk'}\,\mathbf{K}}
\eeq
 where $z=\s+i\tau$. Now we can relate the Riemann theta functions to Jacobi Theta functions (we use the convention in \cite{GR}) by:
 \beq 
 \theta(\zeta) = \theta_3(\pi\zeta,q) , \ \ \ \hat{\theta}(\zeta) = 
    \theta\left[\begin{array}{c} \Delta_1 \\ \Delta_2   \end{array}\right](\zeta) = - \theta_1(\pi\zeta,q), \ \ \ \ \mbox{with} \ \ \ \ q=e^{i\pi\Pi} ,
 \eeq
Some convenient formulas are \cite{GR}
 \beq
   D_1\theta(\hat{a}) = \omega_1 \theta_3'(\frac{1+\Pi}{2}) = i\pi\omega_1 q^{-\frac{1}{4}} \theta'(0)
  \eeq
  and
  \beq
  \theta_3(0) = \sqrt{\frac{2\mathbf{K}}{\pi}}, \ \ \theta'_1(0)=\theta_2(0) \theta_3(0) \theta_4(0) \ \ \  \theta_2(0) \theta_4(0) = \sqrt{kk'} \theta_3^2(0)
  \eeq
 The world-sheet metric (or Lagrange multiplier $\Lambda$) is determined by
  \beq
    e^{2\alpha} = \frac{\theta_3^2(\pi\zeta)}{\theta_1^2(\pi\zeta)} =
     \frac{1}{kk'}\left(\frac{\dn(\frac{2\tau}{\sqrt{kk'}},k)}{\sn(\frac{2\tau}{\sqrt{kk'}},k)}\right)^2
  \eeq
In the last equation $\alpha$ was written in terms of Jacobi elliptic functions, perhaps the easiest expression to check the validity of the cosh-Gordon
equation 
\beq
\frac{1}{4}\partial_\tau^2 \alpha = e^{2\alpha} + e^{-2\alpha}
\eeq
for such function. The last step is to choose the spectral parameter $\lambda$ such that $|\lambda|=1$. In terms of $\lambda$ we define
\beq
v = \inte{1}{4} = -\frac{i}{4 \sqrt{kk'}\ \mathbf{K}} \int_0^\lambda \frac{d\lambda}{\mu(\lambda)}
\eeq
  The solution can then be written in terms of Jacobi theta functions as 
  \beqa
  \mu &=& 2\pi\omega_1 \partial_v \ln \theta_3(v) \\
  \nu &=& \mp 2\pi \omega_1 \partial_v \ln \theta_1(v) \\
  \psi_1 &=& -2\pi\omega_1 \frac{\theta'_1(0) \theta_1(v) \theta_1(x+v)}{\theta_3(0)\theta_3(v)\theta_1(x)} e^{-\frac{\alpha}{2}} e^{\mu z + \nu \bar{z}} \\
  \psi_2 &=& \pm \frac{\theta_3(x+v)}{\theta_3(x)} e^{\frac{\alpha}{2}} e^{\mu z + \nu \bar{z}} \\
  \tilde{\psi}_1 &=& 2\pi\omega_1 \frac{\theta'_1(0) \theta_1(v) \theta_1(x-v)}{\theta_3(0)\theta_3(v)\theta_1(x)} 
    e^{-\frac{\alpha}{2}} e^{-\mu z - \nu \bar{z}} \\
  \tilde{\psi}_2 &=& \pm \frac{\theta_3(x-v)}{\theta_3(x)} e^{\frac{\alpha}{2}} e^{-\mu z - \nu \bar{z}} \\
   \psi_1 \tilde{\psi}_2 - \psi_2 \tilde{\psi}_1 &=& \left|\frac{\theta_1(2v)\theta_3(0)}{\theta_3^2(v)}\right| \\
  \zeta &=& 2 \omega_1 (\pm \bar{z}-z) \\
  \lambda &=& \left(\frac{\theta_1(v)}{\theta_3(v)}\right)^2
  \eeqa
where $v$ is an additional parameter which corresponds to $\inte{1}{4}$ in \eqn{m15}.
The formulas for $\psi_{1,2}$ and $\tilde{\psi}_{1,2}$ refer to the functions defined in \cite{IKZ} and are included for completeness. They are
not used in this paper.  Finally, the solution is written as
\beqa
  Z &=& \left|\frac{\theta_1(2v)\theta_3(0)}{\theta_3^2(v)}  e^{-2\mu z - 2 \nu \bar{z}}\right| 
        \frac{\theta_1(x)\theta_3(x)}{\left|\theta_1(x+v)\right|^2+\left|\theta_3(x+v)\right|^2} \\
  X+iY &=& e^{-2\bar{\mu} \bar{z} - 2 \bar{\nu} z} 
    \frac{\theta_3(x+v)\bar{\theta}_3(x-v)-\theta_1(x+v)\bar{\theta}_1(x-v)}{\left|\theta_1(x+v)\right|^2+\left|\theta_3(x+v)\right|^2}
\eeqa
  As before the boundary curves are found by finding the zeros of $\that=-\theta_1$. Since the variable $\zeta_+$ depends only on $\tau$ 
( or $\zeta_-$ on $\s$), the world-sheet region is a strip between two zeros of $\theta_1$. Generically the edges of the strip map to two
infinite boundary curves with the shape of a spiral. However if we take $\lambda=1$ for $a>1$ and $\lambda=-1$ for $a<1$ we have
the following interesting result
\begin{itemize}
  \item $a>1$, $\lambda=1$, the Wilson loop is given by two concentric circles of radii $r_{1,2}$, satisfying 
   \beq
   \ln \frac{r_2}{r_1}= \pi \mathrm{Re} \left(\frac{\theta_3'(v)}{\theta_3(v)}+\frac{\theta_1'(v)}{\theta_1(v)}\right)
   \eeq
The ratio, by definition, is smaller than one, but, analyzing the previous result, it shows that it has a minimum value larger than zero. 
 For each value of the ratio $\frac{r_2}{r_1}$ in that range there are two possible values of $v$. However they do not give the same solution,
 namely, for each allowed ratio  $\frac{r_2}{r_1}$ there are two solutions, one world-sheet is close to the boundary and the other extends
 towards the interior. 
  \item $a<1$, $\lambda=-1$ the Wilson loop is given by a cusp. 
\end{itemize}
The following limits are of interest
  \begin{itemize}
   \item $a\rightarrow \infty$ gives a circle.
   \item $a\rightarrow 1^{+}$ gives two parallel lines.
   \item $a\rightarrow 1^{-}$ gives two parallel lines.
   \item $a\rightarrow 0$ gives a straight line. 
  \end{itemize}
  For the case $a=1$ corresponding to two parallel line the spectral curve coincides with the one given in \cite{Janik}:
  \beq
  \mu^2 = \lambda (\lambda^2-1)
  \label{m53}
  \eeq
  suggesting that both approaches are equivalent.

\section{Surfaces ending on two Contours}
 
\begin{figure}[h!]
   \centering
     \includegraphics[width=\textwidth]{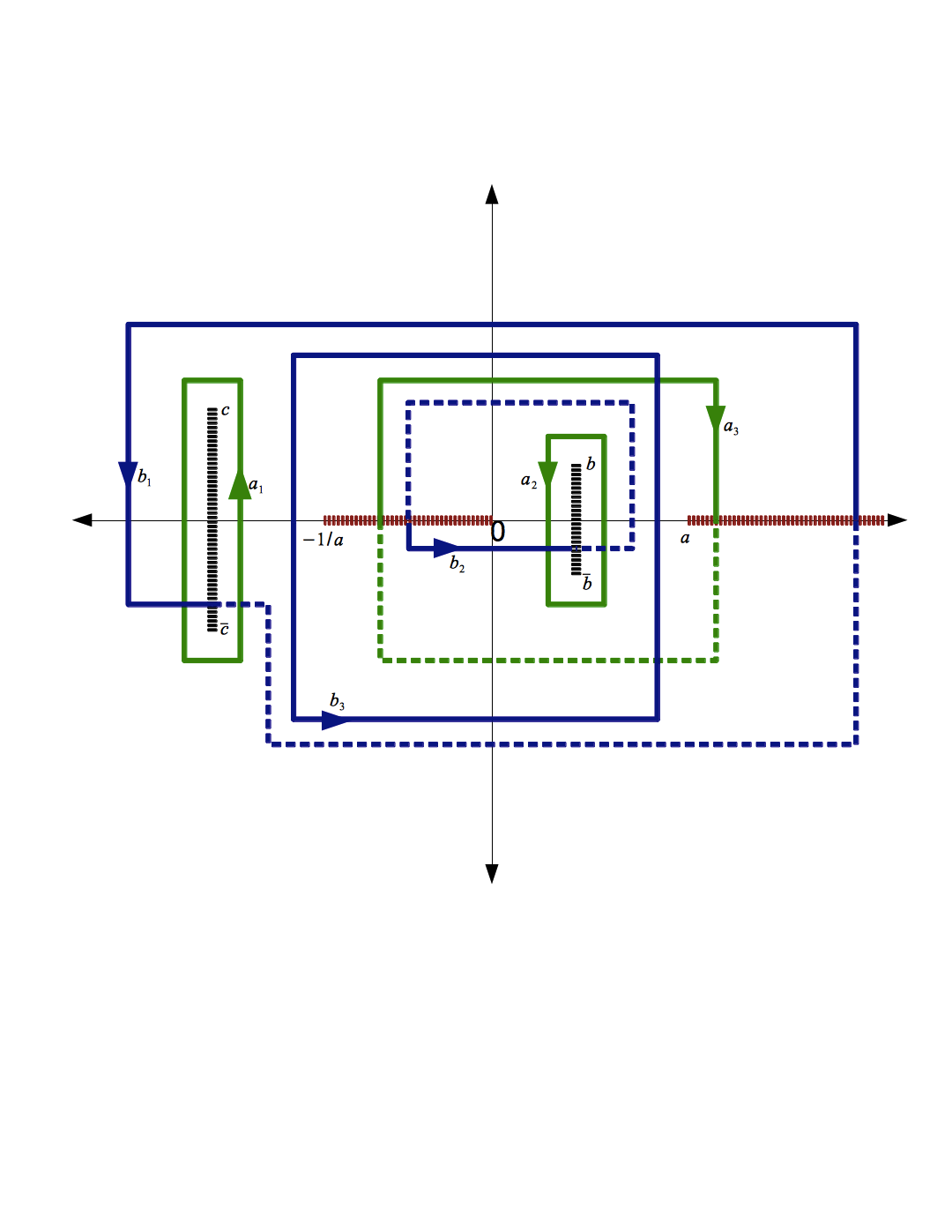}
     \vspace{-1.4in}
      \caption{\em g=3 hyperelliptic Riemann surface.}
      \label{fig3}
 \end{figure}

  In the previous section we discussed, following  \cite{DF} a solution ending on two concentric circles and corresponding to two cuts.
It seems evident that adding two small cuts to the spectral curve will just slightly deform the solution. This is correct except that generically the periodicity is broken and therefore the curves become infinite spirals instead of two closed curves. To be precise, the theta functions are 
periodic but there is also the exponential factor in \eqn{m21} which is periodic when the exponent is purely imaginary, a condition that needs 
to be satisfied. Both periods have to match. It is clear that for particular positions of the cuts the periodicity would be restored but it requires some numerical effort to find them. 
 In this section we present several examples. 
  The number of branch cuts is four, the situation is depicted in fig.\ref{fig3} where also the choice of cycles is made evident. 
Due to the involution $\lambda \to -1/\bar{\lambda}$\,,  the branch point $c=-1/b$. The points $\bar{b}$ and $\bar{c}$ are the complex conjugates of $b$ and $c$, respectively. 
 
 %{\center \includegraphics[width=4.0in,height=4.0in]{hypg3.png}}\\
 %%%%%%%%%%%%%%%%%%%%%%%%%%%%%%%%%%%%%%% 
 
 %%%%%%%%%%%%%%%%%%%%%%%%%%%%%%%%%%%%%%%%%%
  As we shrink the $b\bar{b}$ and $c\bar{c}$ branch cuts we should get results close to the concentric circles for $a>1$.  
Unlike the $g=1$ case, even if we take $\lambda=1$, the solutions are not automatically periodic; we must pick a suitable 
value of $a>1$ as we move $b$ to get periodic concentric curves.
 
 The ratio of Riemann theta functions in \eqn{m21} is periodic if the argument shifts by an integer vector, also, the exponential function is periodic
under a $2\pi$ shift in the imaginary direction. Both periods have to match. This can be achieved in practice by moving the branch points around. 
 We found values for $a$ and $b$ such that we can match 2,3,4 and 5 periods of the theta function into a single period of the exponential. 
 In each case we obtained concentric Wilson loops. The curves deviate from a circle in a way such that the periodicity is manifest as can
be seen in figs.\ref{mc1} and \ref{mc2}.  The corresponding positions of the branch points is shown in table \ref{tab:tab1} below. 
 %%%%%%%%%%%%%%%%%%%%%%%%%%%%%%%%%%%%%%%%%%%%%%%%%%%%%%%%%%
\begin{table}[position specifier]
 \centering
 \begin{tabular}{ | l | l | l | l | l |}
 \hline
  & n=2 & n=3 & n=4 & n=5 \\ \hline
         & $a=1.28088$ & $a=1.102149$ & $a=1.035312$ & $a=1.0304752$ \\  \hline
      & $b=0.5+0.01 i$ & $b=0.5+0.1 i$ & $b=0.7+0.2  i $ & $b=0.7+0.3  i$ \\  \hline
         \end{tabular}
 \caption{\em Positions of the branch points; n corresponds to the number of periods of the exponential prefactor in each period of the ratio of theta function.}
 \label{tab:tab1}
 \end{table}
 %%%%%%%%%%%%%%%%%%%%%%%%%%%%%%%%%%%%%%%%%%%%%%%%%%%%%%%%%%%%%%%%%%%%%
\begin{figure}
    \begin{center}
    \includegraphics[width=0.9 \textwidth]{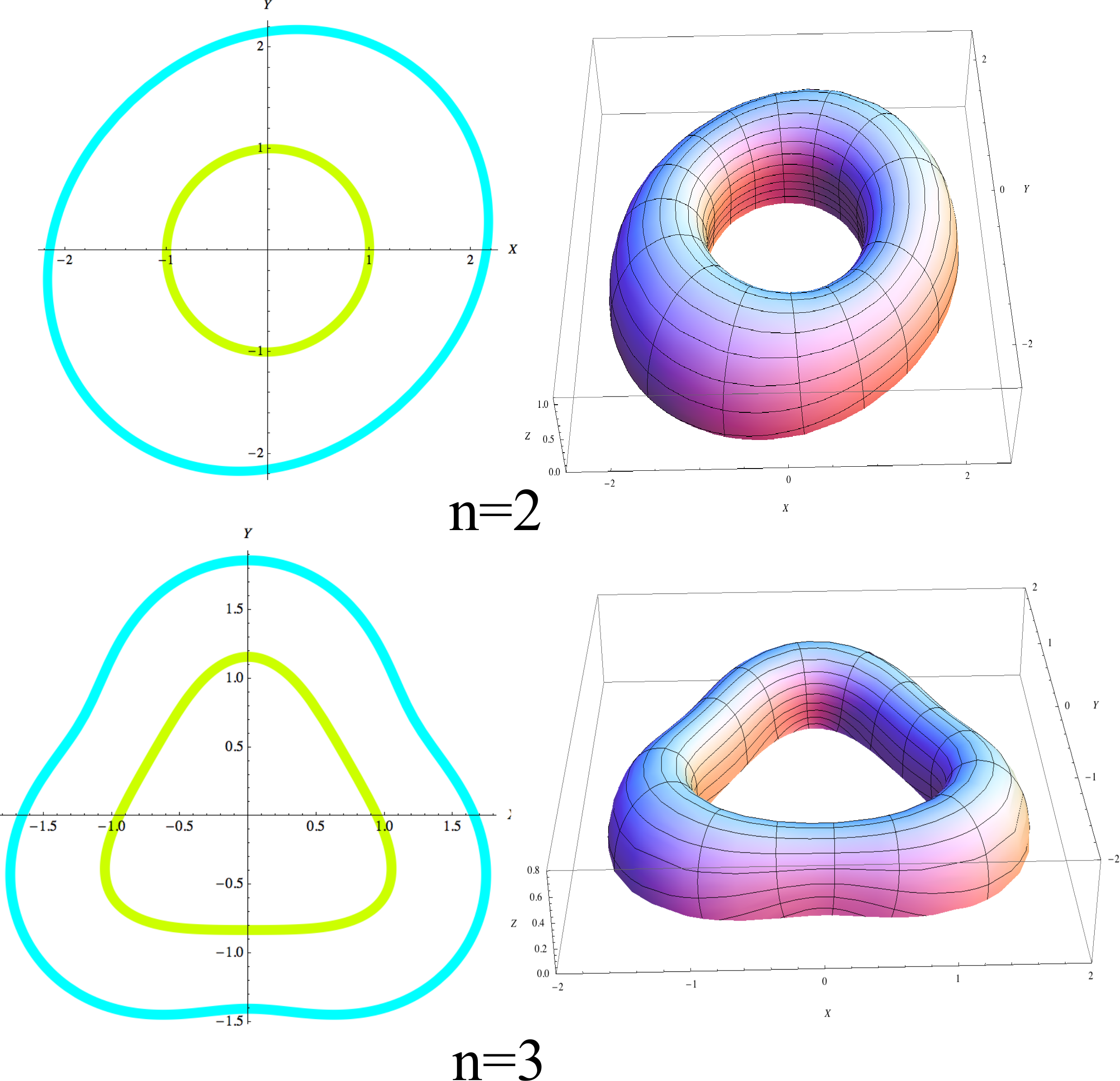} 
    \end{center}
\caption{Wilson loops and corresponding minimal area surfaces for two concentric curves. In this case, two and three periods of the theta functions are fit
into one period of the exponential as is evident from the shape of the curves.}
\label{mc1}
\end{figure}
\begin{figure}
   \begin{center}
   \includegraphics[width=0.9 \textwidth ]{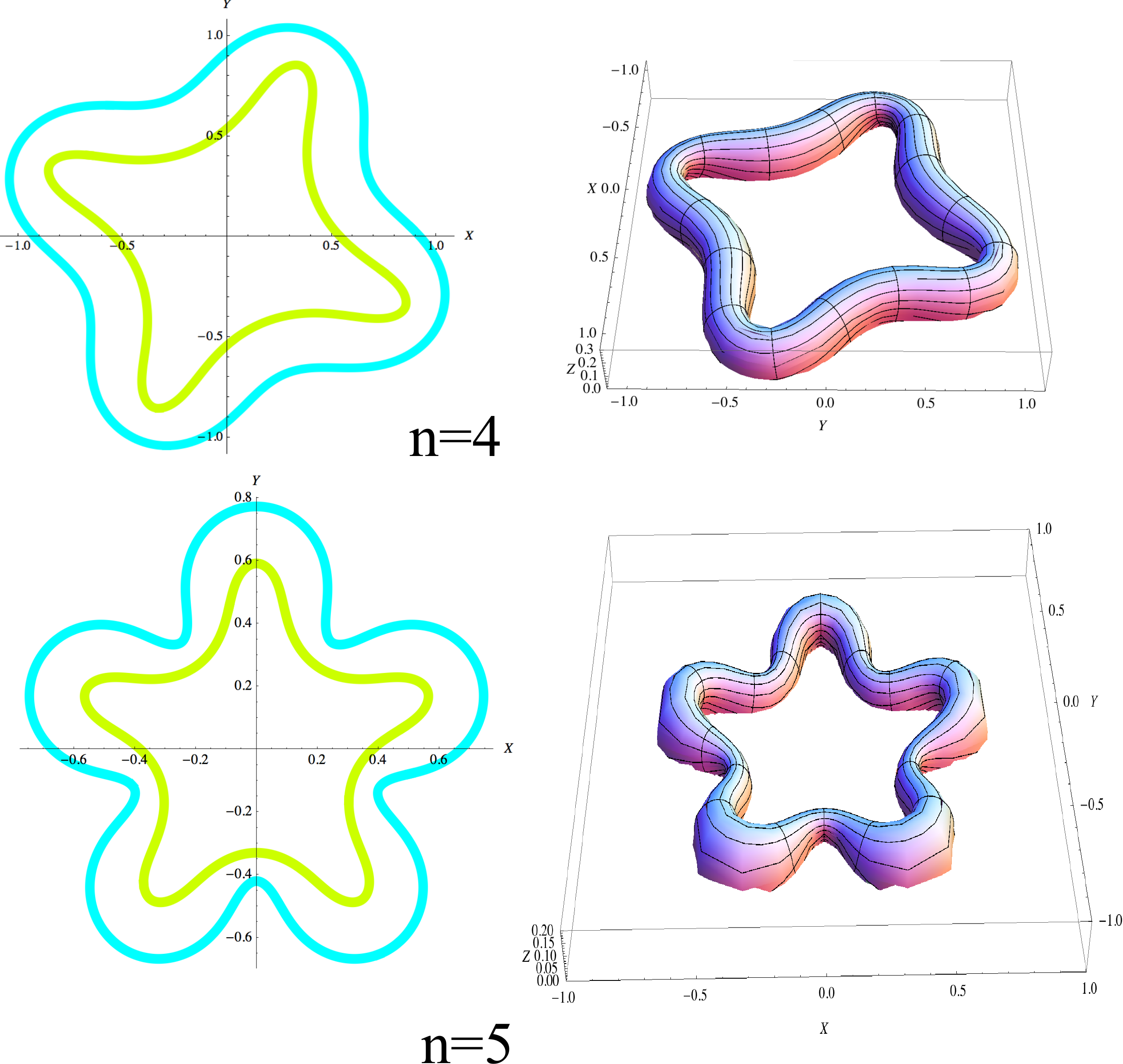} 
\end{center}
\caption{Wilson loops and corresponding minimal area surfaces for two concentric curves. Now, four and five periods of the theta functions are fit
into one period of the exponential as is evident from the shape of the curves.}
\label{mc2}
\end{figure}
 %%%%%%%%%%%%%%%%%%%%%%%%%%%%%%%%%%%%%%%%%%%%%%%%%%%%%%%%%%%%%%%%%%%%%%%%%%%%%%%%%%%%%%%%%%%%%%%%%%%%%%%%%%%%%%%%%%%%%%%%%%%%%%%%%%%%%%%%%%%%%%%%%%%%%%%%%%%%%%%%%%%%%%%%%%%%%%%%%%%%%%%%%%%%%%%%%%%%%%%%%%%%%%%%%%%%%%%%%%%%%%%%%%%%%%%%%%%%%%%%%%%%%%%%%%%%%%%%%%%%
 It is also possible to fit one period of the theta function into \eg\ two periods of the exponential in (\ref{m21}). 
It turns out that we get a boundary made of self-intersecting concentric curves. For appropriate choices of $a$ and $b$ each 
individual curve resembles a circle which is described twice as shown in fig.\ref{mc3}.
\begin{figure}
\begin{center}
   \includegraphics[width=0.9 \textwidth ]{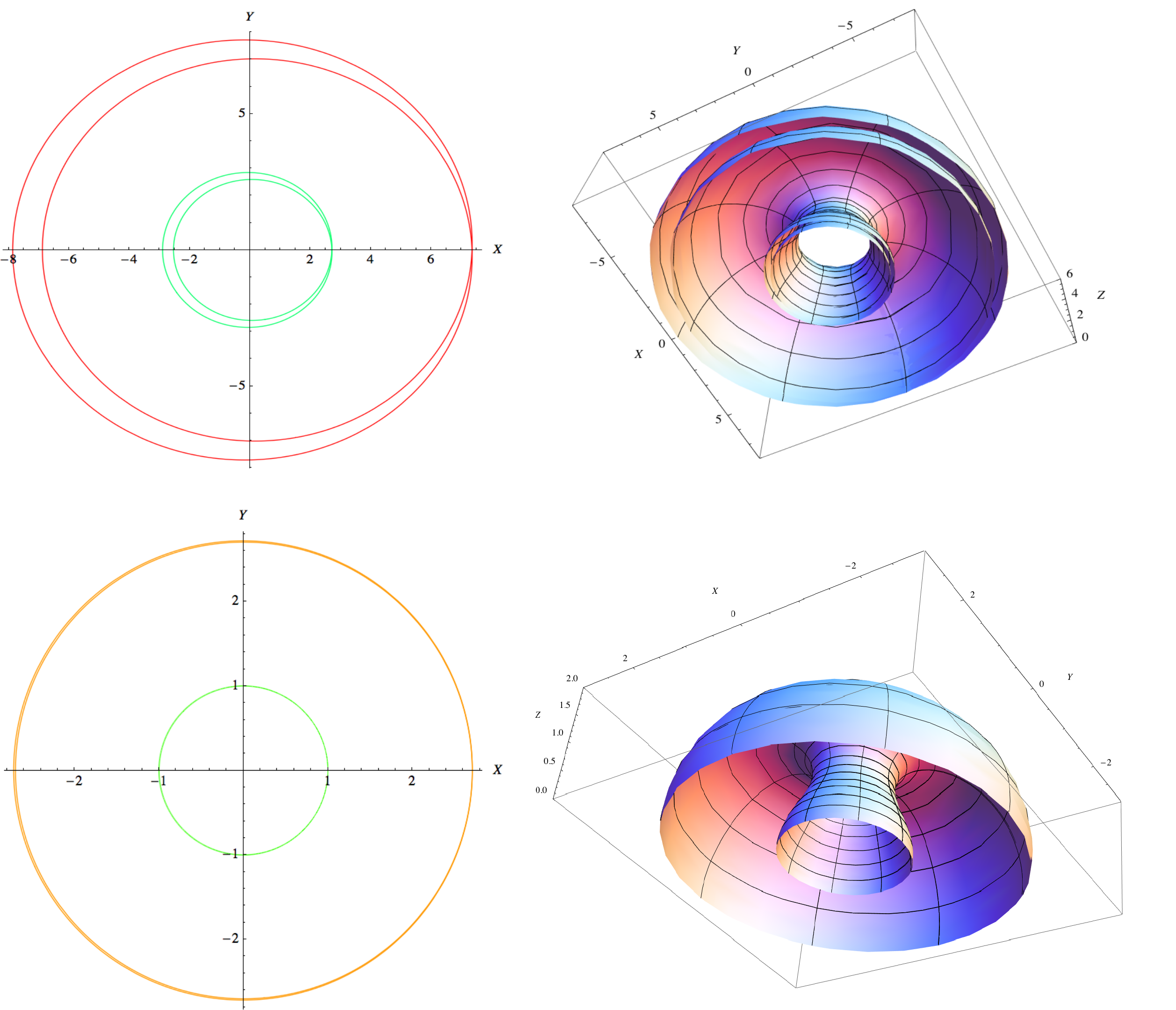} 
\end{center}
\caption{In this case, two periods of the exponential are fit in two periods of the theta function giving rise to concentric self-intersecting curves.
In the first example the position of the cuts is given by $a=2.412712,b=2 + 0.05 I$ and in the second $a=2.414205,b=2 + 0.003 I$. The second one is 
shown as illustration of the fact that the surface can be made to approximate the (double cover) of the concentric circles.}
\label{mc3}
\end{figure}
 It is also made clear the claim that as we shrink the $b\bar{b}$ branch cut the surface approaches the one corresponding to the concentric circular 
Wilson loops. 
%
%  \begin{center}
%   \includegraphics[width=2.5in,height=2.5in]{p1c.png}  \hspace{0.5in} \includegraphics[width=2.5in,height=2.7in]{p2c.png}\\
%   \vspace{-0.01in}
%   {\em Wilson loop}. $n=1/2, a=2.4142,b=2 + 0.005 I$ \hspace{1in} {\em Dual surface}
%   \end{center}
% 

\section{More than two Boundary Curves}
 
\begin{figure}
\begin{center}
     \includegraphics[width=.9\textwidth]{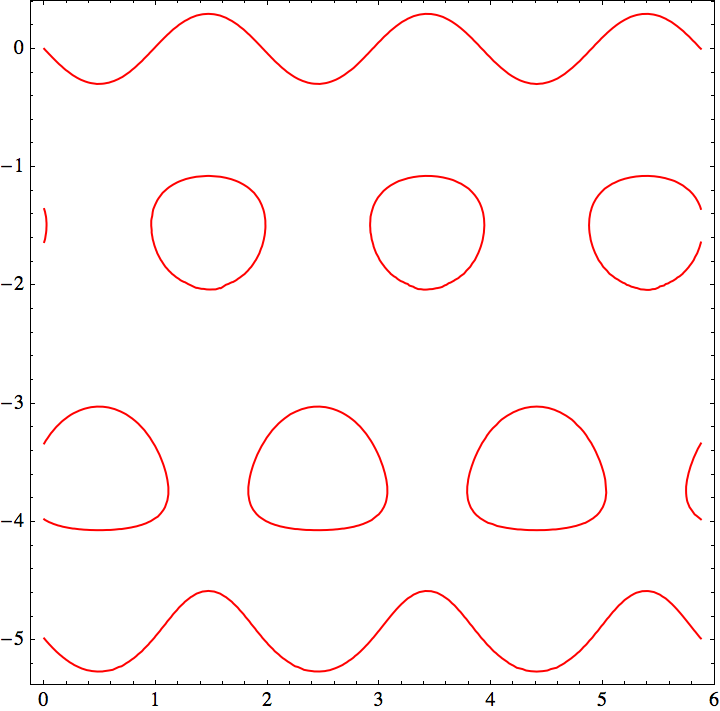}
     \vspace{0.1in}
      \caption{World-sheet of the string where the zeros of $\that$ are depicted. Those curves map to the boundary. For particular position of the 
cuts, the period of the figure in the $\s$ direction matches the period of the exponential function and a surface ending in several closed curves
 is obtained.}
 \label{figSeven}
\end{center}
\end{figure}

\begin{figure}
   \centering
     \includegraphics[width=1.0\textwidth]{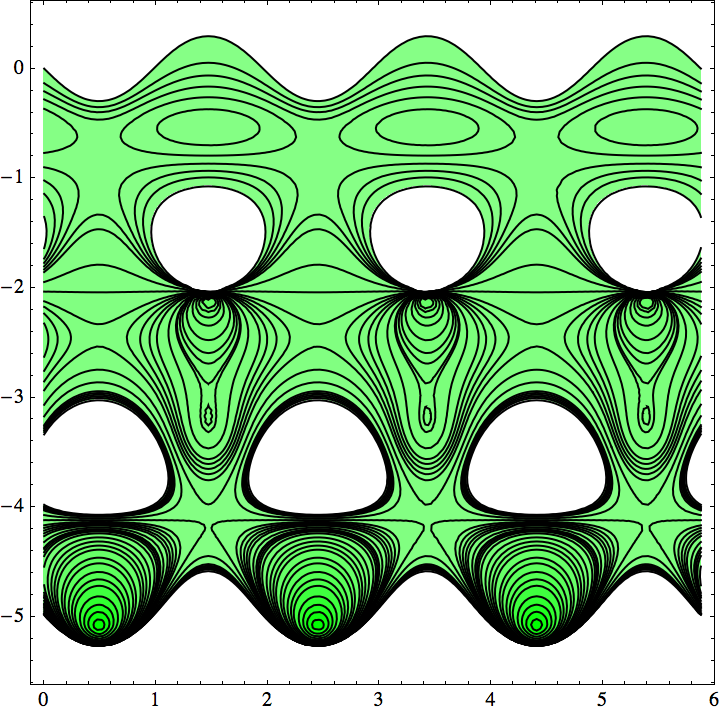}
     \vspace{0.1in}
      \caption{Contour lines indicating the surface heights above the boundary Wilson loop.}
      \label{fig8}
 \end{figure}

\begin{figure}
   \centering
     \includegraphics[width=0.4\textwidth]{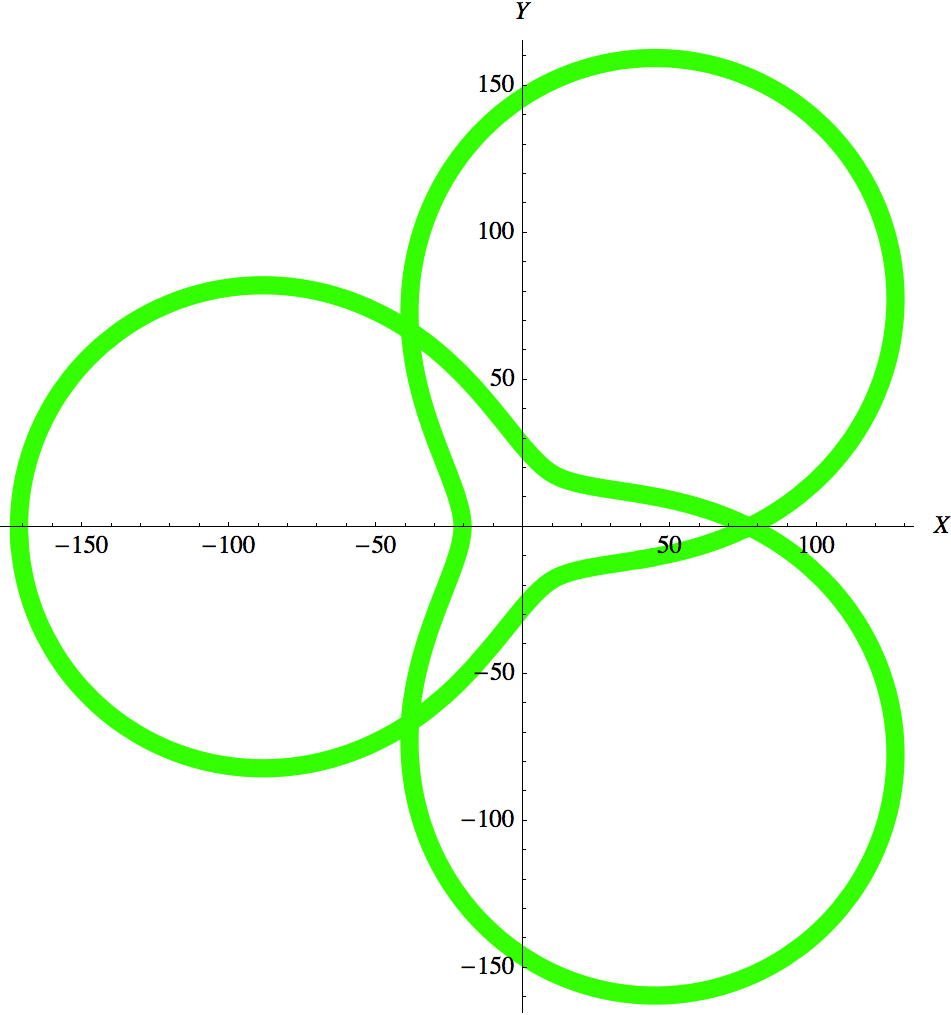} \hspace{0.1in} \includegraphics[width=0.4\textwidth]{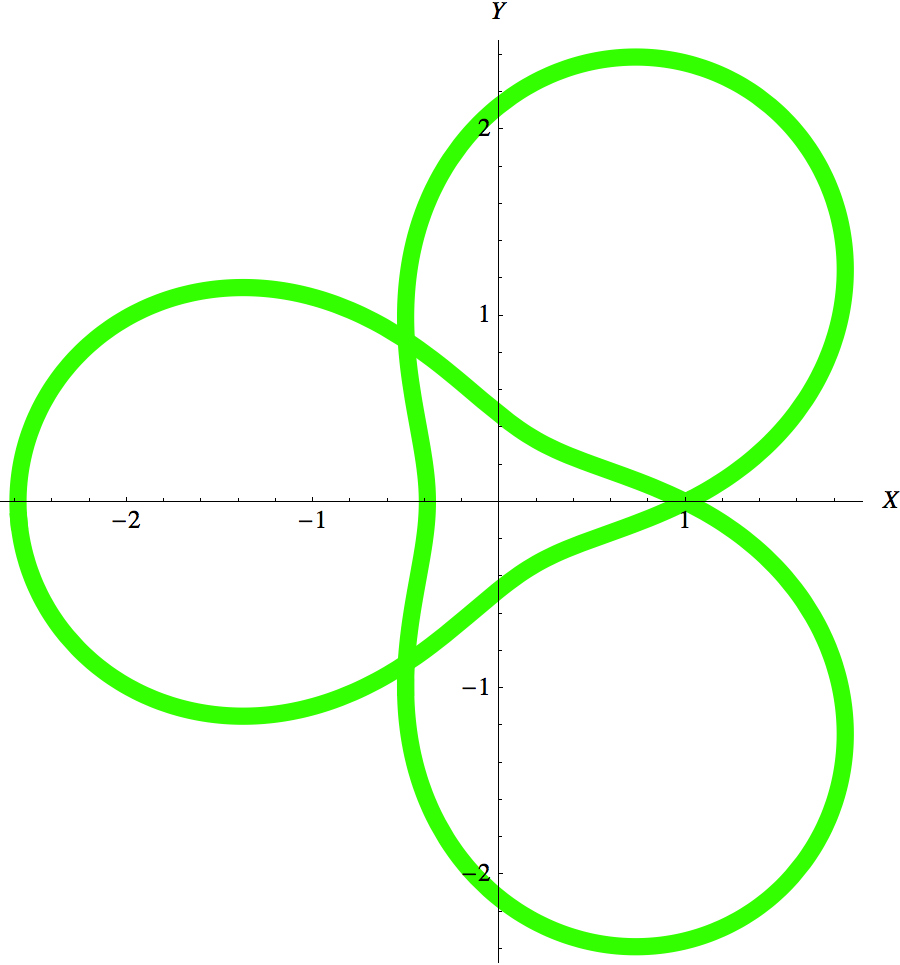}
     \vspace{0.1in}
      \caption{Image on the boundary of the curves depicted in fig.\ref{figSeven}.}
      \label{fig9}
 \end{figure}

\begin{figure}
   \centering
     \includegraphics[width=1.0\textwidth]{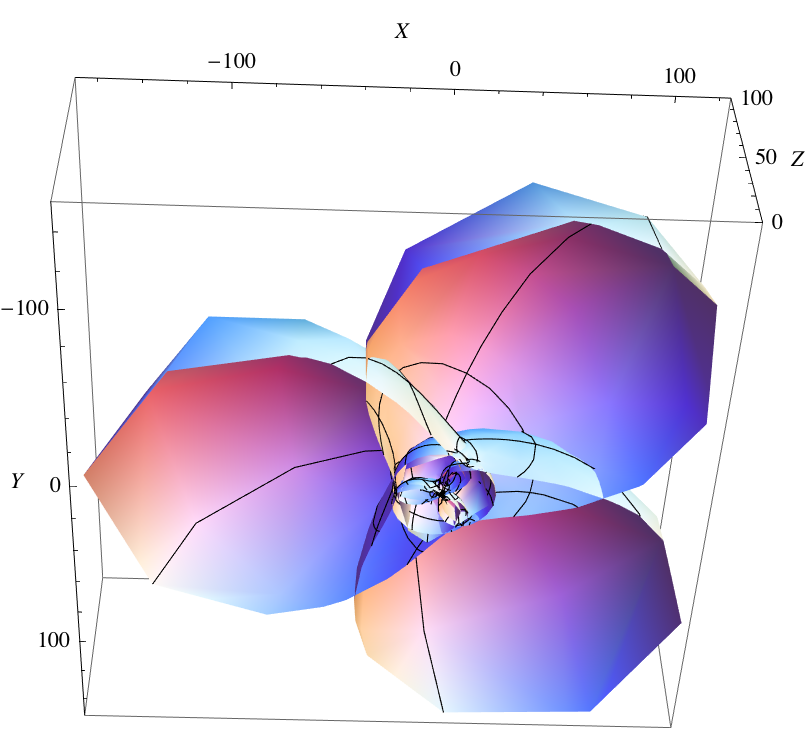} \hspace{0.1in} 
     \vspace{0.1in}
      \caption{Example of minimal area surface embedded in Euclidean $AdS_3$ and ending on multiple contours.}
      \label{fig10}
\end{figure}

 As the size of the cuts is varied, closed curves appear in the region between the two periodic curves giving rise to surfaces with more boundaries. 
An example is shown in fig.\ref{figSeven}. The curves depicted are zeros of $\that$ in the world-sheet and correspondingly of the function $Z(z,\bz)$. That
means that they map into multiple curves at the boundary. The region surrounded by one of the closed curves maps into a minimal area surface 
ending in the corresponding single curve. On the other hand the complement of those regions maps to a minimal area surface ending in multiple curves, 
those closed curves we described plus two infinite curves (of spiral shape) given by the upper and lower periodic curves. Again, for $\lambda=1$ and
particular positions of the cuts, the period of those curves match those of the exponential and the result is a surface ending in multiple closed
curves. 
  A particular example is illustrated in figs. \ref{fig8}, \ref{fig9} and \ref{fig10}. Admittedly, the surface is not simple to describe or plot since it ends
in eight different and self-intersecting curves. The boundary curves are shown in fig.\ref{fig9}, notice however that different scales are used. 
As an intermediate representation, contours of equal $Z$ are depicted in fig.\ref{fig8}. Finally, the surface is plotted in fig.\ref{fig10}. 
 
 %%%%%%%%%%%%%%%%%%%%%%%%%%%%%%%%%%%%%%%%%%%%%%%%%%%%%%%%%%%%%%%%%%%%%%%%%%%%%%%%%%%%%%%%%%%%%%%%%%%%%%%%%%%%%%%%%%%%%%%%%%%%%%%%%%%%%%%%%%%%%%%%%%%%%%%%%%%%%%%%%%%%%%%%%%%%%%%%%%%%%%%%%%%%%%%%%%%%%%%%%%%%%%%%%%%%%%%%%%%%%%%%%%%%%%%%%%%%%%%%%%%%%%%%%%%%%%%%%%%%
 
 \section{Area}

 Consider again the surfaces ending on two concentric curves, namely those in fig.\ref{mc1}, \ref{mc2} and compute their area. The formula 
in \eqn{m49} applies to this case if we take the turning number of the curve to be $n=0$. Remember that this is the turning number of the 
boundary curve in the world-sheet. In fig.\ref{figSeven} it is clear that, as one moves along the upper and lower curves, the tangent to the curves
does not wrap around the unit circle. The area of the world-sheet can also be explicitly written term of a parameterization $\tau_{1,2}(\s)$
of the upper and lower curves giving
\beqa
 A_f&=&4D_{p_1p_3} \log \theta(0) \int{(\tau_2(\sigma)-\tau_1(\sigma))d\sigma}+\frac{1}{2}\left( \int_{\tau_2(\sigma)}+\int_{\tau_1(\sigma)} \right)
 {\frac{\nabla^2\hat{\theta}}{|\nabla \theta|}\,dl}\,  \nonumber \\
 &=&-2 \Im \left\{ D_{13}\log\theta(0) \oint z d\bar{z}+ \left( \int_{\tau_2(\sigma)}-\int_{\tau_1(\sigma)} \right) D_1 \log\theta(\zeta) d\bar{z}\right\}
 \label{eqafinite}
\eeqa 
Recall also that the total area is given by
\beq
 A_{\mbox{total}} = \frac{L}{\epsilon} + A_f
 \label{eqA0}
\eeq
 where $L$ is the length of the Wilson loop and $\epsilon\rightarrow 0$ is the regulator defined by cutting the surface at $Z=\epsilon$.
 
 %%%%%%%%%%%%%%%%%%%%%%%%%%%%%%%%%%%%%%%%%%%%%%%%%%%%%
 \begin{center}
   \begin{tabular}{| c | c | c | c | c |}
     \hline
     Areas & n=2 & n=3 & n=4 & n=5 \\ \hline
     $A_{total}$ by fit \rule[-0.5cm]{0pt}{1.2cm} & $-13.80 + \frac{19.8644}{\epsilon}$ & $-20.55 + \frac{17.25}{\epsilon}$ & $-40.23 + \frac{12.23}{\epsilon}$ & $-55.48 + \frac{9.64}{\epsilon}$  \\ \hline
     $A_{total}$ by eq.(\ref{eqA0}) \rule[-0.5cm]{0pt}{1.2cm} & $-13.80 + \frac{19.86}{\epsilon}$ & $-20.55 + \frac{17.25}{\epsilon}$ & $-40.27 + \frac{12.23}{\epsilon} $ & $-55.66 + \frac{9.63658}{\epsilon}$ \\  \hline
     $A_f $ by eq.(\ref{m48}) \rule[-0.5cm]{0pt}{1.2cm}& $-13.80$ & $-20.55$ & $-40.27$ & $-55.66$ \\ \hline
      \end{tabular}
      \end{center}
 %%%%%%%%%%%%%%%%%%%%%%%%%%%%%%%%%%%%%%%%%%%%%%%%%%%%%%%%%%%%%%%%%%%%%%%%%%%%%%%%%%%%%%%%%%%%%%%%%%%%%%%%%%%%%%%%%%%%%%%%%%%%%%%%%%%%%%%%%%%%%%%%%%%%%%%%%%%%%%%%%%%%%%%%%%%%%%%%%%%%%%%%%%%%%%%%%%%%%%%%%%%%%%%%%%%%%%%%%%%%%%%%%%%%%%%%%%%%%%%%%%%%%%%%%%%%%%

 \section{Conclusions}
 
 In this paper we have simplified and extended the results in \cite{IKZ}. Given a boundary curve $\X(s)$, a one complex parameter family of
curves $\X(s,\lambda)$ can be given by solving the linear problem associated with the flat connection. Studying the analytical properties of
$\X(s,\lambda)$ it turns put that the function is naturally defined on a hyperelliptic Riemann surface $\cM$, the spectral curve. 
 We then denote it as $\X(s,p_4)$ with $p_4\in \cM$. In that way it plays a similar role to the monodromy matrix for standard integrable systems. 
 The first part of the paper is devoted to the study of such function. In particular we find a simple expression for the Schwarzian derivative
$\{\X,s\}$. It is also shown that $\X(s,p_4)$ has an essential singularity at $p_4\rightarrow 0$ and the behavior near the singularity determines 
the area of the surface. 

 The second part of the paper is devoted to the study of surfaces ending in multiple contours. In particular examples for surfaces ending in 
two or more contours are given, generalizing previous results of Drukker and Fiol \cite{DF}. Again, a relatively simple expression is given 
for the area of such curves.

\section{Acknowledgments}

 The motivation for pursuing this research is the result of several discussions  with 
N. Drukker, J. Maldacena, A. Sever, A. Tseytlin and P. Vieira. 
 In addition, M.K. is grateful to V. Enolskii and the other participants of the {\em Algebro-geometric Methods in Fundamental Physics}, 
515 WE-Heraeus Seminar for helpful insights into the mathematical aspects of this work.  
 
 The authors want to thank the Simons Center for Geometry and Physics and M.K. to Perimeter Institute for hospitality while 
part of this work was being done.

 The work of M.K. was supported in part by NSF through a CAREER Award PHY-0952630 and by DOE through grant \protect{DE-SC0007884}
and that of S.Z. in part by the Lyman T. Johnson Postdoctoral Fellowship.

\section{Appendix}

 In this appendix we derive several identities for the Riemann theta functions that are needed in the main text. Some
of the calculations are done in detail to illustrate the procedure. We use a basic knowledge of the theory of theta 
functions as can be found, for example, in the classic book on Riemann surfaces by Farkas and Kra \cite{FK}. In fact 
everything follows from the quasi periodicity property and the trisecant identity. 
Specialized books on theta functions such as \cite{ThF} contain many of these results, we just briefly rederive them in this 
particular context and using the notation of this paper. 

\subsection{Quasi periodicity of $\theta$ and $\that$.}
 
Given two integer vectors $\epsilon_{1,2}$ it follows from the definition of $\theta(\zeta)$ and $\that(\zeta)$, \ie\ \eqns{m7}{m11} that
\beq
 \theta(\zeta+\epsilon_2+\Pi\epsilon_1) = e^{-2\pi i[\epsilon_1^t\zeta+\half\epsilon_1^t\Pi\epsilon_1]} \theta(\zeta)
 \label{b1}
\eeq 
and
\beq
 \that(\zeta+\epsilon_2+\Pi\epsilon_1) = 
 e^{-2\pi i[\epsilon_1^t\zeta+\half\epsilon_1^t\Pi\epsilon_1]}e^{i\pi[\Delta_1^t\epsilon_2-\epsilon_1^t\Delta_2]} \that(\zeta)
  \label{b2}
\eeq
 As discussed in \eqn{m9}, the function $\phi(p_4)=\inte{1}{4}$ is holomorphic on $\cM$ but defined only up to integer periods. The quasi periodicity 
of the theta function allows to write holomorphic functions on $\cM$ as ratios
\beq
 f(p_4) = \frac{\prod_{v=1}^n\theta(\zeta_v+\inte{1}{4})}{\prod_{\tilde{v}=1}^n\theta(\tilde{\zeta}_{\tilde{v}}+\inte{1}{4})}, 
 \ \ \ \ \mbox{with} \ \ \ \ \sum_{v=1}^n \zeta_v = \sum_{\tilde{v}=1}^n \tilde{\zeta}_{\tilde{v}} 
  \label{b3}
\eeq 
for arbitrary constant vectors $\zeta_{v=1\cdots n}$, $\tilde{\zeta}_{\tilde{v}=1\cdots n}$. Indeed, under such condition $f(p_4)$ is independent 
of the path chosen in $\inte{1}{4}$ and therefore is a well defined function on the Riemann surface. A particular useful example is the generalized cross ratio defined in \eqn{b5} below.

\subsection{Trisecant identity}
 
 By studying the analytic properties of ratios such as those in \eqn{b2} Fay derived the following fundamental identity known as trisecant identity \cite{FK}:
\beq
\theta(\zeta) \theta(\zeta+\inte{j}{i}+\inte{k}{l}) = \gamma_{ijkl}\theta(\zeta+\inte{j}{i})\theta(\zeta+\inte{k}{l}) +
  \gamma_{ikjl} \theta(\zeta+\inte{k}{i})\theta(\zeta+\inte{j}{l}) 
   \label{b4}
\eeq  
valid for an arbitrary vector $\zeta\in\mathbb{C}$ and four arbitrary points $p_{i,j,k,l}$ on the Riemann surface. The coefficient $\gamma_{ijkl}$ is 
called a generalized crossed ratio and defined as
\beq
 \gamma_{ijkl} = \frac{\theta(a+\inte{k}{i})\theta(a+\inte{l}{j})}{\theta(a+\inte{l}{i})\theta(a+\inte{k}{j})}
  \label{b5}
\eeq
 where, $a$ is a non-singular zero of the theta function \ie\ $\theta(a)=0$ but $|\nabla \theta|\neq 0$. Under such conditions it is an important 
 property of the cross ratio that it does not depend on the zero $a$ chosen to defined it. In this paper we defined $\that$, see \eqn{m11}, 
through the odd period $\inte{1}{3}$. Since $\that(0)=0$ it implies that $\theta(\inte{1}{3})=0$. Setting $a=\inte{1}{3}$ in the cross ration gives
\beq
\gamma_{ijkl} = \frac{\that(\inte{k}{i})\that(\inte{l}{j})}{\that(\inte{l}{i})\that(\inte{k}{j})}
\eeq
 The trisecant identity plays a central role in using the theta function to solve non-linear differential equations in integrable systems. For us,
it is the starting point for deriving all the identities below. It is interesting to note that in string theory it has an interpretation as a bosonization 
identity \cite{bosonization}.

\subsection{Generalized cross ratio}

 As it was just stated, $\gamma_{ijkl}$ in \eqn{b5} is independent of $a$ as long as $a$ is a zero of the theta function ($\theta(a)=0$). 
In the case at hand, the world-sheet boundary curve $z(s)$ defines a set of zeros of $\that$, namely $\zs$, for any $s$ is a zero of 
$\that$ (see \eqns{m17}{m18}). 
However, from the definition of $\that$ \ie\ \eqn{m11} it is clear that $\that(\zeta_s)=0 \iff \theta(\zs+\inte{1}{3})=0$, it follows that we can take 
 \beq
 a = \zs + \inte{1}{3}
  \label{b6}
 \eeq
  Now choose the points $p_i=p_4$, $p_k=p_3$, $p_l=p_1$, $p_j=\bar{4}$. Remember that $p_1=0$ and $p_3=\infty$. Also $p_4$ and $p_{\bar{4}}$ are the 
same (arbitrary) point in the complex plane but on opposite sheets of the Riemann surface, implying $\inte{1}{4}=-\inte{1}{\bar{4}}$ since $p_1=0$ 
is a branch cut. Up to a constant independent of $\zs$, the cross ratio can be written as 
 \beq
 \tilde{\gamma} = \frac{\theta(\zs+\inte{1}{4})}{\theta(\zs+\inte{1}{\bar{4}})} \frac{\that(\zs+\inte{1}{\bar{4}})}{\that(\zs+\inte{1}{4})} 
  \label{b7}
\eeq
which is therefore independent of $\zs$ as long as it is a zero of $\that$. However, it is also true that $\that(0)=0$. Replacing $\zs \rightarrow 0$ 
and using that $\theta$ is symmetric whereas $\that$ is antisymmetric it immediately follows that $\tilde{\gamma}=-1$. Therefore the simple but very useful 
 identity
 \beq
 \frac{\theta(\zs+\inte{1}{4})}{\theta(\zs-\inte{1}{4})} = - \frac{\that(\zs+\inte{1}{4})}{\that(\zs-\inte{1}{4})} 
 \ \ \ \ \mathbf{if} \ \ \ \that(\zs)=0
  \label{b8}
 \eeq
 is derived.
 
\subsection{First Derivatives}
 
 Identities for the derivatives of the theta functions can be obtained from the trisecant identity by deriving with respect to the position of
the points $p_i$ $p_j$, $p_k$, $p_l$. Consider deriving the trisecant identity with respect of the position of $p_i$ and afterwards taking the limit 
$p_j\rightarrow p_i$. The result reads
\beq
 D_i\ln\frac{\theta(\zeta)}{\theta(\zeta+\inte{k}{l})} = D_i\ln\frac{\that(\inte{l}{i})}{\that(\inte{k}{i})} 
 -\frac{\that(\inte{l}{k})D_i\that(0)}{\that(\inte{i}{k})\that(\inte{l}{i})}
  \frac{\theta(\zeta+\inte{i}{l})\theta(\zeta+\inte{k}{i})}{\theta(\zeta+\inte{k}{l})\theta(\zeta)}
  \label{b8b}
\eeq
 valid for any vector $\zeta\in\mathbb{C}^g$ and any three pints $p_{i}$, $p_{k}$, $p_l$ in $\cM$. 
 In our case, such points are usually chosen among $p_1=0$, $p_3=\infty$, and $p_4$, $p_{\bar{4}}$ arbitrary 
and in opposite sheets of $\cM$. As particular cases it is useful to derive
\beqa
 D_3\ln\frac{\theta(\zeta)}{\theta(\zeta+\inte{1}{4})} &=& -D_3\ln \theta(\inte{1}{4}) 
 -\frac{\that(\inte{1}{4})D_3\that(0)}{\theta(\inte{1}{4})\theta(0)} \frac{\that(\zeta)\that(\zeta+\inte{1}{4})}{\theta(\zeta)\theta(\zeta+\inte{1}{4})}
 \non \\
 D_3\ln\frac{\that(\zeta)}{\that(\zeta+\inte{1}{4})}   &=& -D_3\ln \theta(\inte{1}{4}) 
 +\frac{\that(\inte{1}{4})D_3\that(0)}{\theta(\inte{1}{4})\theta(0)} \frac{\theta(\zeta)\theta(\zeta+\inte{1}{4})}{\that(\zeta)\that(\zeta+\inte{1}{4})}  
 \non \\
 D_1\ln\frac{\theta(\zeta)}{\that(\zeta+\inte{1}{4})}  &=& -D_1\ln \that(\inte{1}{4})  
 +\frac{\theta(\inte{1}{4})D_1\that(0)}{\that(\inte{1}{4})\theta(0)} \frac{\that(\zeta)\theta(\zeta+\inte{1}{4})}{\theta(\zeta)\that(\zeta+\inte{1}{4})} 
 \non \\
 D_1\ln\frac{\that(\zeta)}{\theta(\zeta+\inte{1}{4})}  &=& -D_1\ln \that(\inte{1}{4})  
 +\frac{\theta(\inte{1}{4})D_1\that(0)}{\that(\inte{1}{4})\theta(0)} \frac{\theta(\zeta)\that(\zeta+\inte{1}{4})}{\that(\zeta)\theta(\zeta+\inte{1}{4})}  
 \non\\
  \label{b10}
 \eeqa
 valid for any vector $\zeta$. The function $\that$ is defined in \eqn{m11} and is odd, a fact used in writing \eqn{b10}.  
 In the solutions \eqn{m15}, the vector $\zeta$ is a function of the world-sheet coordinates $(z,\bz)$:
\beq
 \zeta(z,\bz) = 2 \omega_1 \bz + 2\omega_3 z
  \label{b11}
\eeq
and therefore
\beq
 \partial_z F(\zeta(z,\bz)) = 2 D_1 F(\zeta), \ \ \ \partial_{\bz} F(\zeta(z,\bz)) = 2 D_3 F(\zeta)
  \label{b12}
\eeq
In that way, the identities (\ref{b10}) can be used as a table of derivatives of the theta functions and allow the verification 
of the equations of motion. 

\subsection{Second derivatives}

 Second derivatives can be computed in the same way as first ones. In this subsection we just derive a special case that is useful for our purposes.
Consider \eqn{b8b} and take the points $p_i=p_3$, $p_l=p_4$ and use the definition of $\that$, \ie\ \eqn{m11}. After reordering the terms, the result is
\beq
 D_3\ln\frac{\theta(\zeta)}{\that(\zeta+\inte{1}{4}-\inte{3}{k})} = D_3\ln\frac{\that(\inte{3}{k})}{\theta(\inte{1}{4})}
 - \frac{D_3\that(0)\that(\zeta+\inte{1}{4})}{\theta(\inte{1}{4})\theta(\zeta)} 
   \frac{\theta(\zeta-\inte{3}{k})\theta(\inte{1}{4}-\inte{3}{k})}{\that(\inte{3}{k})\that(\zeta+\inte{1}{4}-\inte{3}{k})}
\label{b13c}
\eeq
 The pint $p_k$ is now taken to be very close to $p_3$ so that the expansion in \eqn{m13}
can be used, namely
\beq
 \inte{3}{k} = \omega_3 \tilde{y} + \frac{1}{3} \tilde{y}^3 \omega_{32} +\ldots
 \label{b14c}
\eeq
 with $\tilde{y}=\frac{2}{\sqrt{\lambda}}$. Multiplying \eqn{b13c} by $\that(\inte{3}{k})$ and expanding in powers of $\tilde{y}$ gives, at order 
$\tilde{y}^2$,
\beq
 D_3^2\ln\left[\theta(\zeta)\theta(\inte{1}{4})\that(\zeta+\inte{1}{4})\right] + 
  \left[D_3\ln\frac{\theta(\zeta)\theta(\inte{1}{4})}{\that(\zeta+\inte{1}{4})}\right]^2 = \frac{D_3^3\that(0)}{D_3\that(0)}
  \label{b15c}
\eeq
 Similarly, taking $p_i=p_1$ and expanding fro $p_k\rightarrow p_1$ it follows that
\beq
  D_1^2\ln\left[\theta(\zeta)\that(\inte{1}{4})\theta(\zeta+\inte{1}{4})\right] + 
   \left[D_1\ln\frac{\theta(\zeta)\that(\inte{1}{4})}{\theta(\zeta+\inte{1}{4})}\right]^2 = \frac{D_1^3\that(0)}{D_1\that(0)}
   \label{b16c}
\eeq

\subsection{First derivatives at $\zeta=\zs$.}

 A special role in the construction is played by the world-sheet boundary curve $z(s)$ defined such that
\beq
 \that(\zs)=0
\eeq
where
\beq
 \zs = \zeta(z(s),\bz(s)) =  2 \omega_1 \bz(s) + 2\omega_3 z(s)
  \label{b13}
\eeq
It therefore becomes necessary to evaluate various derivatives at such values of $\zeta=\zs$. 

Consider the first and third equations in (\ref{b10}) and evaluate them at $\zeta=\zs$. The result is
\beqa
 D_3 \ln\frac{\theta(\zs)}{\theta(\zs+\inte{1}{4})} &=& -D_3 \ln \theta(\inte{1}{4})  \label{b14a}\\
 D_1 \ln\frac{\theta(\zs)}{\that(\zs+\inte{1}{4})}  &=& -D_1 \ln \that(\inte{1}{4})   \label{b14b}
\eeqa 
Furthermore, using that $\that$ is antisymmetric and $\theta$ symmetric it follows that
\beqa
 D_3 \ln\frac{\theta(\zs+\inte{1}{4})}{\theta(\zs-\inte{1}{4})} &=& 2 D_3 \ln \theta(\inte{1}{4})  \label{b15a}\\
 D_1 \ln\frac{\that(\zs+\inte{1}{4})}{\that(\zs-\inte{1}{4})}   &=& 2 D_1 \ln \that(\inte{1}{4})   \label{b15b}
\eeqa
Consider now the second equation in (\ref{b10}). Both sides have poles when $\that(\zeta)=0$ but after multiplying it by $\that(\zeta)$,
it reads
\beq
 D_3\that(\zeta)+\that(\zeta) D_3\ln\frac{\theta(\inte{1}{4})}{\that(\zeta+\inte{1}{4})}  =
 \frac{\that(\inte{1}{4})D_3\that(0)}{\theta(\inte{1}{4})\theta(0)} \frac{\theta(\zeta)\theta(\zeta+\inte{1}{4})}{\that(\zeta+\inte{1}{4})}
  \label{b16}
\eeq 
 Evaluating at $\zeta=\zs$ gives
\beq
  D_3\that(\zs) =
  \frac{\that(\inte{1}{4}) D_3\that(0)}{\theta(\inte{1}{4})\theta(0)}
  \frac{\theta(\zs)\theta(\zs+\inte{1}{4})}{\that(\zs+\inte{1}{4})}  
   \label{b17}
\eeq
Similarly, from the last equation in (\ref{b10}) it follows that
\beq
  D_1\that(\zs) = 
  \frac{\theta(\inte{1}{4}) D_1\that(0)}{\that(\inte{1}{4}) \theta(0)}
  \frac{\theta(\zs)\that(\zs+\inte{1}{4})}{\theta(\zs+\inte{1}{4})}  
   \label{b18}
\eeq
 Another possibility to cancel the pole at $\that(\zeta)=0$ is by subtracting the second equation in (\ref{b10}) from itself evaluated at $p_{\bar{4}}$:
\beq
    D_3 \ln \frac{\that(\zeta+\inte{1}{4})}{\that(\zeta-\inte{1}{4})}   = 2 D_3 \ln \theta(\inte{1}{4})
    - \frac{\that(\inte{1}{4}) D_3\that(0)}{\theta(\inte{1}{4})\theta(0)}\frac{\theta(\zeta)}{\that(\zeta)}
    \left[\frac{\theta(\zeta+\inte{1}{4})}{\that(\zeta+\inte{1}{4})}
    +\frac{\theta(\zeta-\inte{1}{4})}{\that(\zeta-\inte{1}{4})}\right]
     \label{b22}
\eeq
where again we used that $\that$ is odd: $\that(-\inte{1}{4})=-\that(\inte{1}{4})$. 
From the trisecant identity, taking the points $p_i=p_4$, $p_j=p_1$, $p_k=p_3$, $p_l=p_{\bar{4}}$ and using 
that $\inte{1}{\bar{4}}=-\inte{1}{4}$ one obtains
\beq
\left[\frac{\theta(\zeta+\inte{1}{4})}{\that(\zeta+\inte{1}{4})}
    +\frac{\theta(\zeta-\inte{1}{4})}{\that(\zeta-\inte{1}{4})}\right] =
     \frac{\that(2\inte{1}{4})\theta(0)}{\that(\inte{1}{4})\theta(\inte{1}{4})} \frac{\theta(\zeta)\that(\zeta)}{\that(\zeta+\inte{1}{4})\that(\zeta-\inte{1}{4})} 
 \label{b23}
\eeq
 Replacing in \eqn{b22}, $\that(\zeta)$ indeed cancels resulting in
\beq
     D_3 \ln \frac{\that(\zeta+\inte{1}{4})}{\that(\zeta-\inte{1}{4})}   = 2 D_3 \ln \theta(\inte{1}{4})
     - \frac{\that(2\inte{1}{4}) D_3\that(0)}{\theta^2(\inte{1}{4})}
    \frac{\theta^2(\zeta)}{\that(\zeta+\inte{1}{4})\that(\zeta-\inte{1}{4})} 
      \label{b24}
\eeq
 which is now safe to evaluate at $\zeta=\zs$. Similarly, from the last equation in \eqn{b10} it follows that
\beq
    D_1 \ln \frac{\theta(\zeta+\inte{1}{4})}{\theta(\zeta-\inte{1}{4})}   = 2 D_1 \ln \that(\inte{1}{4})
    - \frac{\theta(\inte{1}{4}) D_1\that(0)}{\that(\inte{1}{4})\theta(0)}\frac{\theta(\zeta)}{\that(\zeta)}
    \left[\frac{\that(\zeta+\inte{1}{4})}{\theta(\zeta+\inte{1}{4})}
    +\frac{\that(\zeta-\inte{1}{4})}{\theta(\zeta-\inte{1}{4})}\right]
     \label{b25}
\eeq 
leading to
\beq
     D_1 \ln \frac{\theta(\zeta+\inte{1}{4})}{\theta(\zeta-\inte{1}{4})}   = 2 D_1 \ln \that(\inte{1}{4})
     - \frac{\that(2\inte{1}{4}) D_1\that(0)}{\theta^2(\inte{1}{4})}
    \frac{\theta^2(\zeta)}{\theta(\zeta+\inte{1}{4})\theta(\zeta-\inte{1}{4})} 
      \label{b26}
\eeq

\subsection{Higher derivatives at $\zeta=\zs$.}

 Consider \eqn{b14a}.and take $p_4\rightarrow p_3$. Both sides develop a pole so we should carefully expand using \eqn{m13} 
\beqa
 \omega(p_4) &=& \omega_3 + \tilde{y}^2 \omega_{32}  + \ldots \\  
 \inte{1}{4} \omega &=& \inte{1}{3} \omega +\omega_3 \tilde{y} +\frac{1}{3}\tilde{y}^3\omega_{32} + \ldots \\
  \label{b27}
\eeqa
where $\tilde{y} = \frac{2}{\sqrt{\lambda}} $. It follows that
\beq
\frac{D_3\that( \omega_3 \tilde{y}+\frac{1}{3} \tilde{y}^3\omega_{32}+\ldots )}{\that( \omega_3 \tilde{y}+\frac{1}{3} \tilde{y}^3\omega_{32}+\ldots  )} = 
 \frac{D_3\that(\zs+  \omega_3 \tilde{y}+\frac{1}{3} \tilde{y}^3\omega_{32}+\ldots   )}
 {\that(\zs+ \omega_3 \tilde{y}+\frac{1}{3} \tilde{y}^3\omega_{32}+\ldots  )} - D_3\ln\theta(\zs)
  \label{b28}
\eeq
 Expanding both sides in $\tilde{y}$ two useful identities are derived:
\beqa
 \frac{D_3^2\that(\zs)}{D_3\that(\zs)}&=& 2 D_3\ln \theta(\zs) \label{b29a} \\
 \frac{D_3^3\that(\zs)}{D_3\that(\zs)}  &=&\frac{D_3^3\that(0)}{D_3\that(0)}-\frac{D''_3\that(0)}{D_3\that(0)}   
  + \frac{D''_3\that(\zs)}{D_3\that(\zs)} 
    + 3 \left(D_3\ln\theta(\zs)\right)^2 \label{b29b}
\eeqa
where $D''_3$ denotes the derivative along the direction $\omega_{32}$ and we also took into account that $\that$ is an odd function and 
therefore its even derivatives vanish at $0$. Similarly, starting from \eqn{b14b} one finds
\beqa
  \frac{D_1^2\that(\zs)}{D_1\that(\zs)}&=& 2 D_1\ln \theta(\zs)  \label{b30a} \\
  \frac{D_1^3\that(\zs)}{D_1\that(\zs)}  &=&\frac{D_1^3\that(0)}{D_1\that(0)}-\frac{D''_1\that(0)}{D_1\that(0)}   
   + \frac{D''_1\that(\zs)}{D_1\that(\zs)} 
     + 3 \left(D_1\ln\theta(\zs)\right)^2  \label{b30b}
\eeqa
 Finally, taking two derivatives in direction $\omega_3$ \ie\ applying $D_3^2$ to \eqn{b16} and evaluating at $\zeta=\zs$ follows that
\beq
 D_3^2 \ln\left[\theta(\zs)\theta(\zs+\inte{1}{4})\that(\zs+\inte{1}{4})\right]
+\left(D_3\ln\frac{\theta(\zs+\inte{1}{4})}{\that(\zs+\inte{1}{4})}\right)^2 =
 \frac{D_3^3\that(\zs)}{D_3\that(\zs)} - 3 \left(D_3\ln\theta(\zs)\right)^2
  \label{b31}
\eeq
where \eqns{b14a}{b17} and \eqn{b24} were used to simplify the result. Similarly
\beq
 D_1^2 \ln\left[\theta(\zs)\theta(\zs+\inte{1}{4})\that(\zs+\inte{1}{4})\right]
+\left(D_1\ln\frac{\theta(\zs+\inte{1}{4})}{\that(\zs+\inte{1}{4})}\right)^2 =
 \frac{D_1^3\that(\zs)}{D_1\that(\zs)} - 3 \left(D_1\ln\theta(\zs)\right)^2
  \label{b32}
\eeq
is obtained. Instead for this derivation one can simply replace $\zeta=\zs$ in \eqns{b15c}{b16c}. The result is not exactly the same, they only agree if
\beq
\frac{D_3^3\that(\zs)}{D_3(\zs)}-\frac{D_3^2\theta(\zs)}{\theta(\zs)} - 2\left(D_3\ln\theta(\zs)\right)^2 = 
 \frac{D_3^3\that(0)}{D_3\that(0)}-\frac{D_3^2\theta(0)}{\theta(0)}
 \label{b32c}
\eeq
 which is another useful identity. 

 \subsection{Other identities}
 
 In \cite{IKZ} the following result was obtained
\beqa
 \left(\frac{\theta(0)\that(\inte{1}{4})}{D_1\that(0)\theta(\inte{1}{4})}\right)^2 &=& -4\lambda_{p_4} \label{b33a}\\
 \left(\frac{D_3\that(0)\that(\inte{1}{4})}{\theta(0)\theta(\inte{1}{4})}\right)^2 &=& -\frac{1}{4}\lambda_{p_4} \label{b33b}
\eeqa
 Consider now the first equation in \eqn{b10}, apply $D_3$ and evaluate at $\zeta=\zs$:
\beq
 D_3^2 \ln\frac{\theta(\zs)}{\theta(\zs+\inte{1}{4})} = -\left(\frac{\that(\inte{1}{4})D_3\that(0)}{\theta(\inte{1}{4})\theta(0)}\right)^2
 = \frac{1}{4}\lambda_{p_4}
 \label{b34}
\eeq
 where \eqn{b17} and \eqn{b33b} were used to simplify the result. Similarly, from the third equation in (\ref{b10}) and using \eqns{b18}{b33a} 
 it follows that
\beq
 D_1^2 \ln\frac{\theta(\zs)}{\that(\zs+\inte{1}{4})} = \left(\frac{\theta(\inte{1}{4})D_1\that(0)}{\that(\inte{1}{4})\theta(0)}\right)^2
 =-\frac{1}{4\lambda_{p_4}}
 \label{b35}
\eeq
 These identities are valid whenever $\zs$ is a zero of $\that$. In particular we can set $\zs=0$ obtaining
\beqa
 D_1^2 \ln\hat{\theta}(\int_1^4) &=& \frac{1}{4\lambda} + \frac{D_1^2\theta(0)}{\theta(0)} \label{b36a}\\
 D_3^2 \ln{\theta}(\int_1^4) &=& -\frac{\lambda}{4} + \frac{D_3^2\theta(0)}{\theta(0)}  \label{b36b}
\eeqa

\end{document}